\documentclass[pageno]{apaper}

\usepackage[normalem]{ulem}
\usepackage[fleqn]{amsmath}								
\usepackage{amssymb}
\usepackage{amsthm}
\usepackage{graphicx}		
\usepackage{placeins}
\usepackage{sidecap}
\usepackage{listings, multicol}								
\usepackage{tabu}
\usepackage{xspace}
\usepackage{bbm}
\usepackage{float} 									%
\usepackage{algorithm}
\usepackage{xfrac}                                  
\usepackage{pifont}                                 
\usepackage[noend]{algpseudocode}
\usepackage{multirow,booktabs}
\usepackage{multicol}
\usepackage{minibox}
\usepackage{rotating}
\usepackage{setspace}
\usepackage{bm}
\usepackage{tikz}
\usetikzlibrary{fit,shapes, positioning, backgrounds}
\usepackage[dvipsnames]{xcolor}
\usepackage{pifont}
\usepackage{tcolorbox}

\makeatletter
\newcommand*{\transpose}{%
  {\mathpalette\@transpose{}}%
}
\newcommand*{\@transpose}[2]{%
  \raisebox{\depth}{$\m@th#1\intercal$}%
}
\makeatother

\newcommand{\ourmethod}{\textit{BaCO}\xspace} 
\newcommand{\etal}{{\em et al.}\xspace}

\newcommand{\tyes}{{\color{ForestGreen}\ding{51}}\xspace }
\newcommand{\tpartial}{{\color{ForestGreen}\ding{51}}*\xspace }
\newcommand{\tno}{\color{BrickRed}\scalebox{1.2}{\bm{$\times$}}\xspace }

\begin{document}

\title{
BaCO: A Fast and Portable Bayesian Compiler Optimization Framework
}

\date{}

\author{%
\begin{tabular}{p{.33\textwidth} p{.33\textwidth} p{.333\textwidth}} Erik Hellsten{\small$^1$} & Artur Souza{\small$^2$} & Johannes Lenfers{\small$^3$}\\ 
\small{Lund University} & \small{Federal University of Minas Gerais} & \small{University of Münster}\vspace{.2cm}\\
Rubens Lacouture{\small$^4$} & Olivia Hsu{\small$^5$} & Adel Ejjeh{\small$^6$} \\ 
\small{Stanford University} & \small{Stanford University} & \small{University of Illinois UC}\vspace{.2cm}\\
Fredrik Kjolstad{\small$^7$} & Michel Steuwer{\small$^8$} & Kunle Olukotun{\small$^9$} \\ 
\small{Stanford University} & \small{University of Edinburgh} & \small{Stanford University} \vspace{.2cm}\\ 
&Luigi Nardi{\small$^{10}$} & \\ 
&\small{Lund University, Stanford University} &\end{tabular} }

\maketitle
\blfootnote{$^{1,10}$\{erik.hellsten, luigi.nardi\}@cs.lth.se, $^2$arturluis94@gmail.com, $^3$j.le@wwu.de, $^{4,5,7,9}$\{rubensl, owhsu, kjolstad, kunle\}@stanford.edu, $^6$aejjeh@illinois.edu, $^8$michel.steuwer@ed.ac.uk}
\thispagestyle{empty}

\begin{abstract}
\noindent
We introduce the Bayesian Compiler Optimization framework (BaCO), a general purpose autotuner for modern compilers targeting CPUs, GPUs, and FPGAs. BaCO provides the flexibility needed to handle the requirements of modern autotuning tasks. Particularly, it deals with permutation, ordered, and continuous parameter types along with both known and unknown parameter constraints. To reason about these parameter types and efficiently deliver high-quality code, BaCO uses Bayesian optimization algorithms specialized towards the autotuning domain. We demonstrate BaCO's effectiveness on three modern compiler systems: TACO, RISE\,\&\,ELEVATE, and HPVM2FPGA for CPUs, GPUs, and FPGAs respectively. For these domains, BaCO outperforms current state-of-the-art autotuners by delivering on average 1.36$\times$--1.56$\times$ faster code with a tiny search budget, and BaCO is able to reach expert-level performance 2.9$\times$--3.9$\times$ faster.

\end{abstract}

\section{Introduction}
Modern compilers are rapidly evolving to keep pace with the growing range of increasingly specialized hardware targets, as well as the ever-changing domains of interest.
A recent trend is to separate policy (what to compute) from mechanism (transformations and code generation describing how to compute) by using scheduling languages.
Prominent examples of this paradigm include Halide~\cite{ragan2013halide}, TVM~\cite{chen2018tvm}, TACO~\cite{senanayake2020scheduling}, and RISE\,\&\,ELEVATE~\cite{hagedorn2020achieving,steuwer2022rise}.
This design pushes the optimization task of finding good schedules outside of the compiler core, where it can be done manually or automatically by an autoscheduler.
Scheduling languages may express more complex optimization spaces, and, thus, require more advanced autotuning features to effectively and efficiently tackle the autoscheduling task.
Modern hardware backends---like GPUs, as in RISE\,\&\,ELEVATE~\cite{steuwer2022rise}, and FPGAs, as in  HPVM2FPGA~\cite{kotsifakou2018hpvm,ejjeh2022hpvm2fpga}---further increase the complexity of relevant optimization spaces.

\begin{figure}
    \centering
    \includegraphics[width=1\linewidth]{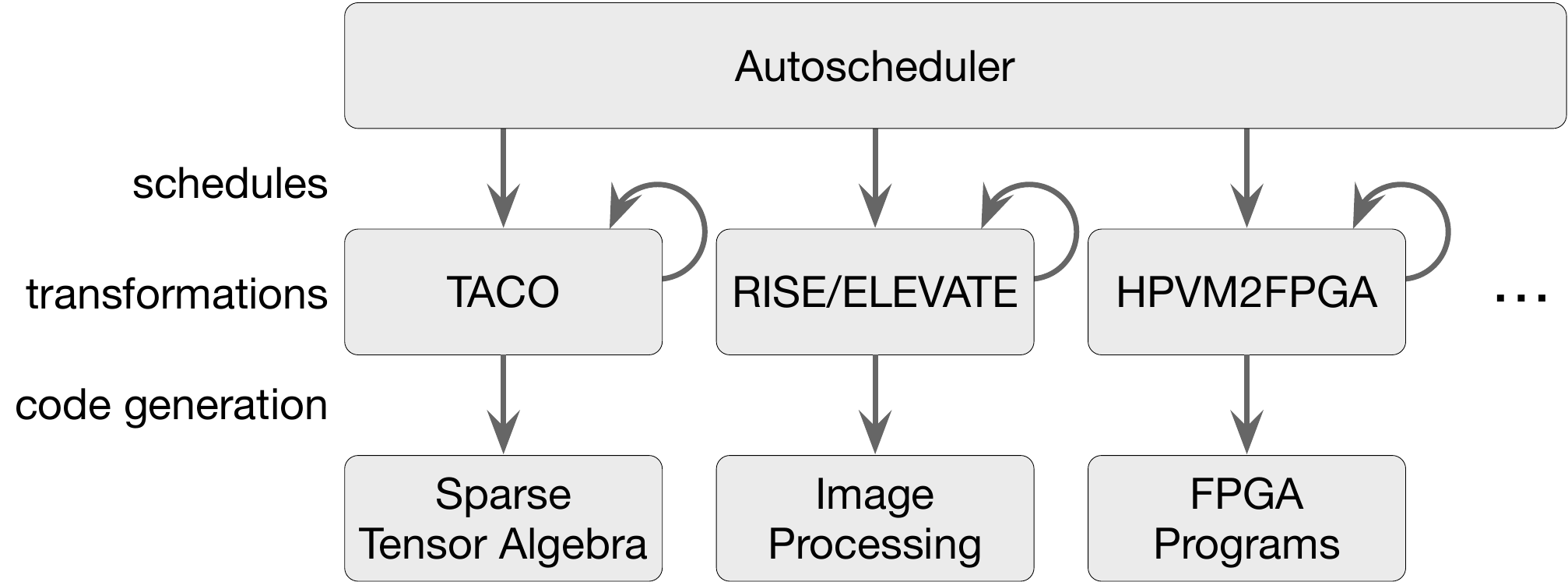}
    \caption{An autoscheduler that is portable across the scheduling languages of diverse compilers.}
    \label{fig:design}
\end{figure}

The separation of concerns between policy and mechanism in compilers exposes a great opportunity. If we can design a portable autoscheduler that is effective across many compilers, like the design shown in Fig.~\ref{fig:design}, then we can reduce the complexity of the overall ecosystem. New compilers get an autoscheduler with minimal effort, and improvements in the autoscheduler automatically benefit all compilers and their subsequent domains and backends.

However, a portable autoscheduler must be designed with a rich input language to allow users to accurately describe the search space exposed by their particular compiler. This autoscheduling search space is determined by the product of the hardware target, the compiler's scheduling language features and configuration tuning parameters. In modern compilers, this search space is often complex, including both continuous parameters (e.g., real-valued tuning parameters), and discontinuous parameters broken down into integers (e.g., tiling factors), permutation categories (e.g., loop reordering), ordinals/ordered categories (e.g., unroll factors), and categoricals/unordered categories (parallelization schemes). These parameter types are often abbreviated to RIPOC~\cite{nardi2019,opentuner14,haj2020autophase}.

However, even the large class of search spaces that can be generated as the Cartesian combination of these parameters is often inadequate. Scheduling parameters frequently depend on the settings of other parameters, leading to constraints on the scheduling space. One such example is a loop bound that must be an exact multiple of a given tiling factor. We refer to these as \textit{known constraints}, which are provided to the autoscheduler ahead of time. Other constraints are initially unknown and must be learned throughout the autoscheduling. An example of this is learning sets of parameters that would generate programs that adhere to hardware constraints, such as avoiding out-of-memory errors on a GPU. Such constraints are often referred to as \textit{hidden constraints}. For a general autotuning framework to be efficient and portable across a multitude of compilers, it needs to support as many of these features as possible.

Once able to express a complex search space, the autoscheduler must be effective and efficient in finding a good schedule within this space.
For optimizing compilers, the performance of the generated code is of primary concern. Therefore, the autoscheduler will invariably use some sort of search combined with a cost model to evaluate points of the search space. The cost model could be analytic or data-driven, but the most accurate cost model is to generate and run the code on its target platform. For a general autoscheduler used across diverse compilers, a cost model based on running the code makes it easy to use the autoscheduler with a new compiler. We refer to empirical autoschedulers, whose cost model is to run the actual code generated by the compiler, as \emph{autotuners}.
To achieve composability, and to work effectively and easily across a diverse set of compilers, it is vital for an autoscheduler to treat each compiler as a black-box system.
The autoscheduler's job will then be to optimize the black-box system using the smallest possible budget of trials and errors, i.e., evaluations of the black-box system. 

Many successful autotuning frameworks have been proposed, some of which are listed in Table~\ref{tab:method_comparison}.
\begin{table}[tb]
    \centering
    \resizebox{\columnwidth}{!}{
    \begin{tabular}{|l|cccc|}
        \hline                                      &    \multicolumn{2}{c}{Parameters} & \multicolumn{2}{c|}{Constraints} \\ \cline{2-5} 
                                                    & RIOC  & Perm. & Hidden          & Known          \\ \hline
        ATF \cite{RaschSSG21}                       & \tyes & \tno  & \tno            & \tyes          \\
        OpenTuner \cite{opentuner14}                & \tyes & \tyes  & \tno           & \tno          \\
        Ytopt \cite{wu2021autotuning}               & \tyes & \tno  & \tno            & \tyes           \\
        Kernel Tuner \cite{willemsen2021}           & \tyes & \tno  & \tno            & \tyes          \\
        KTT \cite{petrovivc2020benchmark}           & \tno  & \tno  & \tno            & \tyes          \\
        GPTune \cite{liu2021gptune}                 & \tyes & \tno  & \tno            & \tyes          \\    
        HyperMapper \cite{nardi2019}                & \tyes & \tno  & \tyes           & \tno           \\
        Bliss \cite{roy2021bliss}                   & \tno  & \tno &  \tno            & \tno            \\
        DeepHyper \cite{dorier2022hpc}              & \tyes & \tno  & \tno            & \tpartial           \\
        SMAC3 \cite{lindauer2022smac3}              & \tyes & \tno  & \tno            & \tpartial           \\
        GpyOpt \cite{gpyopt2016}                    & \tno  & \tno  & \tno            & \tyes          \\
        Spearmint \cite{snoek2012practical}         & \tyes & \tno  & \tyes           & \tno           \\
        GPflowOpt \cite{GPflowOpt2017}              & \tno  & \tno  & \tyes           & \tno           \\
        cBO  \cite{gardner2014bayesian}             & \tno  & \tno  & \tyes           & \tno           \\
        \textbf{BaCO (ours)}                        & \tyes & \tyes & \tyes           & \tyes          \\\hline
    \end{tabular}
    }
    \caption{Framework capabilities; RIOC abbreviates Real/Integer/Ordinal/Categorical parameters. Limited support for constraints is marked with~\tpartial.}
    \label{tab:method_comparison}
\end{table}
These frameworks have helped deliver high-performance software in the past.
However, they do not support all features required to effectively search over the complex search spaces described by diverse scheduling languages across modern domain-specific compilers targeting various hardware backends. For example, Table~\ref{tab:compiler_comparison} shows the features required by three modern compilers (TACO, RISE\,\&\,ELEVATE, and HPVM2FPGA), and Table~\ref{tab:method_comparison} shows that none of 14 recently proposed autotuners support all required features in the manner we define below. Some of the frameworks in Table~\ref{tab:method_comparison} do support some types of user-defined constraints. Several frameworks are using ConfigSpace~\cite{configspace}, which supports conjunctions of linear constraints. While this is certainly useful, it is inadequate for defining non-linear dependencies (see frameworks marked with~\tpartial). The hidden constraints column in Table~\ref{tab:method_comparison} means that the framework uses specialized tools to handle the constraint validity instead of assigning a high objective value to infeasible configurations, such as OpenTuner. Handling hidden constraints with a specialized tool distinguishes slow but valid configurations from invalid configurations, providing more information to the search mechanism. Lastly, while permutations can be cast as categoricals, BaCO is the first autotuning framework to make use of their underlying structure.

\begin{table}[tb]
    \centering
    \resizebox{\columnwidth}{!}{
    \begin{tabular}{|l|cccc|}
        \hline                          &    \multicolumn{2}{c}{Parameters} & \multicolumn{2}{c|}{Constraints} \\ \cline{2-5} 
                                        & RIOC  & Perm. & Hidden          & Known          \\ \hline
        TACO \cite{kjolstad2017taco}    & \tyes & \tyes & \tyes           & \tyes          \\
        RISE\,\&\,ELEVATE \cite{hagedorn2020achieving}            & \tyes & \tyes & \tyes           & \tyes          \\
        HPVM2FPGA \cite{ejjeh2022hpvm2fpga}                    & \tyes &   & \tyes           &            
        \\\hline
    \end{tabular}
    }
    \caption{Features needed by different compilers; 
    RIOC abbreviates Real/Integer/Ordinal/Categorical parameters.}
    \label{tab:compiler_comparison}
\end{table}

Hence, we propose \ourmethod, a novel general autotuning framework optimized towards the autotuning of modern compilers, which efficiently handles all features mentioned above. \ourmethod{} does not require any user-provided cost model but instead learns from observations from running the generated code throughout the optimization procedure. The support for sophisticated search spaces and online learning means that \ourmethod{} finds good schedules in fewer iterations than existing autotuners, while being easy to use. Notably, we do not adapt \ourmethod to individual compilers nor applications but show that it yields expert-level performance out of the box for a wide range of applications.
Our contributions are:
\begin{itemize}
    \item The first Bayesian autotuning framework that supports all RIPOC features, including permutation types, 
    through the definition of separate distance metrics, thus improving search performance
    (Sec.~\ref{sec:framework} and Sec.~\ref{sec:search_spaces}).
    \item The integration of a feasibility model for hidden constraints, simplifying the portability to new compiler backends (Sec.~\ref{sec:search_spaces}).
    \item The first system to use the chain-of-tree technique in a Bayesian optimization setting for portable autotuning on sparse search spaces (Sec.~\ref{sec:search_spaces}).
    \item Applying autotuning to three distinct compiler frameworks for different domains and hardware targets, along with a survey of the autotuning challenges of those three recent domain-specific compilers (Sec.~\ref{sec:complexity}).
\end{itemize}

The end result is that \ourmethod generates expert-level code significantly faster than the state of the art. We demonstrate the effectiveness and robustness of \ourmethod across three real-world compilers and code generation systems targeting CPUs, GPUs, and FPGAs (Sec.~\ref{sec:evaluation}).
\ourmethod reliably produces good schedules across different compilers without any customization for each compiler, such as hyperparameter tuning or custom constraint filtering.
It achieves expert performance on TACO 3.15$\times$--5.0$\times$ faster than the state-of-the-art,  while RISE\,\&\, ELEVATE achieves 1.35$\times$--1.58$\times$ better performance with a tiny autotuning budget, and
HPVM2FPGA achieves peak performance 2.43$\times$--2.77$\times$ faster.

\section{Complexity of Modern Autotuning}
\label{sec:complexity}
To develop the next generation general-purpose autotuning framework, we need to better understand the real-life challenges faced by various modern compiler frameworks.
Therefore, we investigate the autotuning features needed by the TACO, RISE\,\&\,ELEVATE, and HPVM2FPGA compiler frameworks.
This will allow us to identify an autotuning framework that is able to generalize across a wide spectrum of compilers and backend targets. As we shall see, this ideal general-purpose autotuning framework needs to support the features described in Table~\ref{tab:compiler_comparison}, which includes support for a wide range of parameter types and both hidden and known constraints.

\label{sec:compiler_frameworks}
\paragraph{The Tensor Algebra Compiler (TACO).} TACO~\cite{kjolstad2017taco} is the state-of-the-art compiler for sparse tensor algebra.
It generates high-performance code for tensor operations expressed in a high-level Einstein notation, such as the sampled dense-dense matrix multiplication (SDDMM) computation represented as $A(i, j) = B(i, j) * C(i, k) * D(k, j)$.
A particular strength of TACO is its capability to generate code for a large variety of sparse tensor formats~\cite{chou:2018:formats}.

TACO's scheduling language defines an iteration space transformation framework that dictates how to traverse a tensor stored in any particular format~\cite{senanayake2020scheduling}.
This provides a way to introduce optimization transformations, such as tiling, parallelization, vectorization, loop reordering, and more.
An autotuning framework selecting the optimizations exposed by the scheduling language needs to provide not just traditional real, integer, ordinals, and categorical parameters as provided by most frameworks in Table~\ref{tab:method_comparison}, but also \emph{permutation} parameters for selecting loop reordering.
Typically, optimization is performed inside the compiler and controlled by a heuristic, but in TACO, as well as other similar compilers with scheduling languages, it is exposed as a tunable parameter. These optimization parameters also need to follow known constraints that TACO provides. An example is loop reordering variables that TACO enforces for concordant traversal.

\paragraph{RISE\,\&\,ELEVATE.} RISE~\cite{steuwer2022rise} and ELEVATE~\cite{hagedorn2020achieving} are a powerful combination of compiler and scheduling languages.
Computations are described in the RISE~\cite{steuwer2022rise} language using well-known data-parallel patterns like \texttt{map} and \texttt{reduce} in the spirit of LIFT \cite{hagedorn2018liftstencil,steuwer2017lift}.
Optimizations are applied and described in the ELEVATE~\cite{hagedorn2020achieving} scheduling language as compositions of semantic preserving rewrite-rules.
The optimized RISE program is compiled to high-performance CPU or GPU code. 

Transformations, such as loop tiling, may introduce numerical tuning parameters, such as a tile size, which are often constrained by other numerical values, such as loop bounds.
When automatically optimizing RISE programs, an explorative rewrite process speculatively applying program transformations is performed.
To evaluate the performance of a transformed program, the system relies on an autotuning framework to pick all numerical parameters while respecting all known parameter constraints that the system can collect automatically and provide to the autotuning framework.
Compiling for GPUs also introduces hidden constraints for the autotuning framework, such as choosing only parameter values that will result in a program fitting in the tight register and memory requirements.
When these constraints are not satisfied, the compiler generates code that will fail to execute.
Therefore, the autotuning framework must be able to learn these hidden constraints automatically.

\paragraph{HPVM2FPGA.} HPVM2FPGA~\cite{ejjeh2022hpvm2fpga} is a compiler that enables hardware-agnostic programming of field-programmable gate arrays (FPGAs).
The compiler uses sophisticated optimizations, coupled with design space exploration (DSE), to automatically tune and generate well-performing FPGA designs from programs that have not been written by hardware and FPGA experts.
HPVM2FPGA is part of the Heterogeneous Parallel Virtual Machine (HPVM) compiler infrastructure~\cite{kotsifakou2018hpvm,ejjeh2022hpvm}, which provides a retargetable virtual ISA and compiler IR for programming heterogeneous systems.

During HPVM2FPGA's DSE, compiler transformations, such as loop unrolling, greedy loop fusion, argument privatization, and kernel fusion are explored.
HPVM2FPGA generates its parameter space automatically through a static analysis of the IR, and the design space varies depending on the size of the application being compiled.
The majority of the parameters are boolean parameters, with hidden constraints among them, making it challenging to explore the space efficiently.

\section{The BaCO Framework}
\label{sec:framework}

We introduce \textit{Bayesian Compiler Optimization} (BaCO)\footnote{Baco is Italian for bug.}, a Bayesian optimization (BO) framework that learns high-performing auto-scheduling strategies. BaCO thrives in a small data world where configuration evaluations are costly, either due to high runtimes of the kernel or expensive simulations of code generation passes. BaCO is backend-agnostic, and it can be equally applied to CPU, GPU, and FPGA compilers. 
Building on the BO paradigm, BaCO is centred around a \textit{configuration recommendation-evaluation} loop: it recommends promising new configurations that are subsequently scheduled and evaluated by the corresponding compiler toolchain. The evaluation results are used to fit two predictive models: one modelling the predicted value and one modelling the predicted probability of feasibility of new configurations. To initialize the two models, the procedure starts with an \textit{initial phase}, where the first few configurations are sampled uniformly at random from the search space.

\begin{figure}[tb]
    \centering
    \includegraphics[width=.45\textwidth]{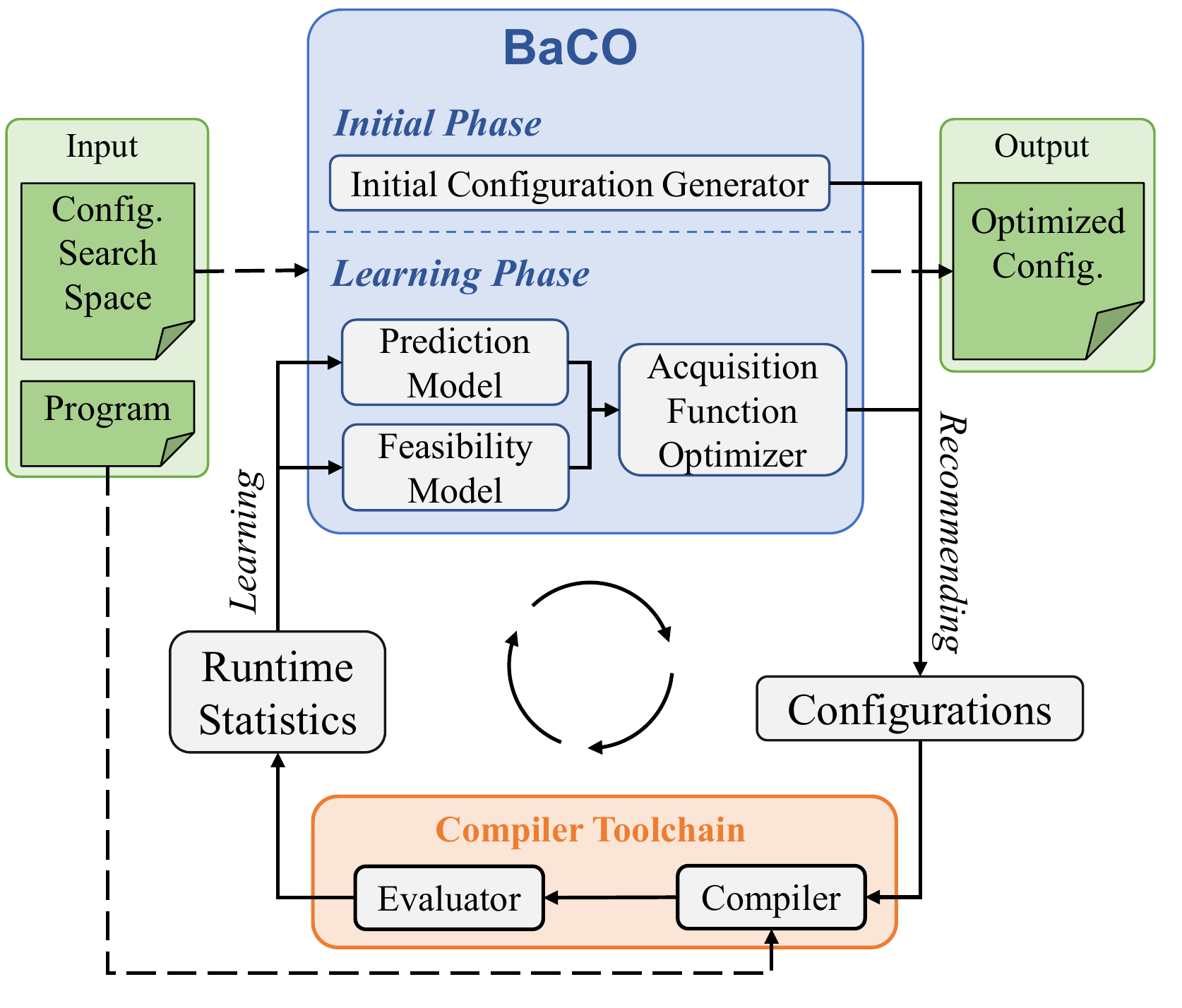}
    \caption{Overview of the BaCO framework.}
    \label{fig:architecture}
\end{figure}

However, for BO to reach its full potential, it needs to be customized for autotuning tasks. 
This section explains the various modules of BaCO's architecture, shown in Fig.~\ref{fig:architecture}, whereas further specialization towards autotuning search spaces is emphasized in Sec.~\ref{sec:search_spaces}.

\subsection{Bayesian optimization}
Bayesian optimization is a steadily growing methodology for solving black-box optimization problems. Those are problems where the objective function $f(\bm{x})$ can only be accessed point-wise through expensive evaluations. At the core of BO is the use of a surrogate model, which estimates the objective function. This helps with performing well-informed decisions about which configurations to evaluate next. The goal is to find a good configuration in as few evaluations as possible. The most common choices of surrogate model are \textit{Gaussian Process} (GP) or \textit{Random Forest} (RF) predictors, which both provide useful uncertainty estimates. For any given point $\bm{x}$, the model provides mean and variance, which is used to balance exploration and exploitation. This trade-off is quantified by an acquisition function. Common examples are \textit{Expected Improvement} and \textit{Lower Confidence Bound}. The surrogate model is dynamically updated to learn from observations, creating a feedback loop where the BO framework proposes new points, that are then evaluated. The information from the evaluation is subsequently used to train the model. BO was historically developed for continuous compact domains, and the extension towards more exotic search spaces is currently being used in this work and studied by the BO community.

\subsection{Surrogate models over compiler domains}
\label{sec:surrogate}
\paragraph{Choice of probabilistic model}
One core element of an efficient BO algorithm is an accurate surrogate model~\cite{frazier2018tutorial}. While complex parameter domains have little impact on less intricate methods such as random sampling, the success of BO depends greatly on clever handling of such parameters. While traditionally, Random Forests have been considered the natural choice as surrogate models over discrete domains~\cite{hutter2011sequential,nardi2019,balaprakash2018autotuning}, recent studies show that a careful implementation of Gaussian Processes yields superior accuracy~\cite{garrido2020dealing, cuesta2022comparison}. 
However, to achieve the true potential of GPs in autotuning and DSE applications, significant customization of the GP is needed. This customization is explained in detail in the following sections. Sec.~\ref{sec:results} shows the impact of some of these major design choices and a comparison between GPs and RFs.

\paragraph{GP kernel similarity function}
A key feature in autotuning and DSE is the mixed-variable search space. Thus, the kernel needs to combine distance measures over different parameter types. We propose the weighted Euclidean norm $||\bm{d}||_2^2 = \sum_{i=1}^D\left(d_i/l_i\right)^2$ over the vector of individual distance measures $\bm{d}$, as a unified distance measure. $D$ denotes the dimension of the search space, i.e., the number of parameters being optimized, and $l_i$ are the horizontal lengthscales, learned using maximum likelihood estimation (MLE)~\cite{murphy2012machine}, weighting the different parameters. We use the 5/2-Matérn kernel~\cite{rasmussen2006}, given by 
\begin{align}
    \label{eq:matern_kernel}
    k(\bm{x},\bm{x}')&= \sigma\left(1 + \sqrt{5}d + 5d^2\right)e^{-\sqrt{5}d},\\
    d &= \sqrt{\sum_{i=1}^D\frac{d(x_i,x_i')^2}{l_i^2}}
    \label{eq:matern_kernel2}
\end{align}
where $d(x_i,x_i')$ denotes the distance between $x_i$ and $x_i'$ (described in Sec.~\ref{sec:distance_metrics}), as this has shown to be efficient in many real life applications~\cite{klein2017fast, torun2018global}. To increase stability, we assume that the value observed in each evaluation, $y(\bm{x})$, is perturbed by some normally distributed noise~\cite{frazier2018tutorial}, such that $y(\bm{x})=f(\bm{x}) + \varepsilon$ and $\varepsilon\sim N(0,\sigma_{\varepsilon})$. 

\paragraph{GP hyperparameter optimization}
A crucial element in effective optimization using GPs is to find good hyperparameters for the model. Especially important are the length-scales $l_i$ presented in Eq.~\eqref{eq:matern_kernel2}, which balance the importance ratio between different parameters. The remaining hyperparameters are the outputscale $\sigma$ in Eq.~\eqref{eq:matern_kernel}, and the magnitude of the Gaussian noise $\sigma_{\varepsilon}$.  BaCO optimizes the hyperparameters using a multistart gradient descent approach, which first uniformly samples a number of possible hyperparameter settings, then chooses a fraction of those with highest likelihood, and optimizes them individually using L-BFGS~\cite{liu1989limited}.

Discrete parameter spaces offer a number of practical challenges when fitting the GP model. One such challenge is that the model hyperparameter optimization method described above frequently prefers to give close to zero lengthscale values to some parameters. In practice, this means that configurations which take different values for those parameters have close to zero similarity, making the GP behave as a sparse model. This is undesirable as it reduces the model's expressive power. To address this artifact of GP modeling, as well as to stabilize the hyperparameter selection, BaCO uses \textit{gamma priors}~\cite{murphy2012machine} for the lengthscales. These priors are chosen to be flexible while cutting out extreme hyperparameter settings. In practice, stabilizing the lengthscales means that different parameters are given more equal importance, preventing certain parameters from becoming too dominant or too insignificant due to model over-fitting. Gamma prior distributions are chosen as they have positive support, can be made reasonably concentrated and have long tails towards both zero and infinity. Other good alternatives with similar properties would be the log-normal or inverse-gamma distributions. By normalizing the input data, BaCO can use a single set of priors that works well for the majority of parameters.
Note that, this is an artifact from that many parameters take identical values in discrete spaces, which rarely occurs when working with continuous parameters.

\subsection{Acquisition function}
\label{sec:acquisition}
The acquisition function quantifies the anticipated utility of evaluating a new point. We use the Expected Improvement (EI) acquisition function~\cite{jones1998efficient}, which balances exploration and exploitation.
Autotuning and DSE are characterized by both discrete search spaces and noisy function evaluations, in which case we observe that standard EI has a tendency to overly prioritize re-sampling points with good values. To avoid this unintended behavior, we propose a modified EI acquisition function which predicts the expected improvement of observing a noise free evaluation of the blackbox function.
Computing the EI without considering the noise in the GP makes sampling repeated points less likely. 

BaCO optimizes the acquisition function by multi-start local search. Initially, a large number of configurations are sampled at uniformly random, of which the best configurations are chosen as starting points for the local search. Neighbours are defined as all configurations that can be reached by modifying a single parameter.

\section{Adapting to Exotic Search Spaces}
\label{sec:search_spaces}

When implementing an efficient autotuning framework, effectively handling all of the search space features is key. As BaCO is built around a GP predictive model, careful design of the distance metrics used for different variable types is of additional importance. In this section we study the intricacies of the different parameter types as well as how to handle known and hidden constraints.
\subsection{Parameter types}
\label{sec:distance_metrics}
\paragraph{Continuous, integer, and ordinal parameters} These types of parameters have the property that the values are comparable, i.e., you can use the greater or equal sign to order them. This can naturally and explicitly be translated into a distance metric, and in particular we use the absolute difference, $d(x_i,x_i')=|x_i-x_i'|$.
However, certain such parameters are innately exponential in nature, such as tile size parameters. 
In that case, we use the Euclidean distance over a log-transformed space instead, $d(x_i,x_i')=|\log{x_i}-\log{x_i}'|$. The log transformation often more accurately describes the relationship between values. Consider tile sizes as an example. We expect the tile sizes 2 and 4 to be roughly as similar to each other as the tile sizes 512 and 1024. However, the tile sizes 512 and 514 would be much more similar than the pairs above.

\paragraph{Categorical parameters} 
Categorical parameters differ from ordinals in that they have no inherent order. Here, we use the \textit{Hamming distance}, defined as $d_{h}(x_i, x_i')=\mathbbm{1}_{x_i\neq x_i'}$, where $\mathbbm{1}$ is the indicator function, which returns $1$ if $x_i\neq x_i'$ and $0$ otherwise. In other words, the Hamming distance only considers whether the parameter values are identical or not. The scale here is not relevant as the distance is weighted by the lengthscale $l_i$ in Eq.~\eqref{eq:matern_kernel2}. 

\begin{figure}[tb]
    \centering
    \includegraphics[width=.45\textwidth]{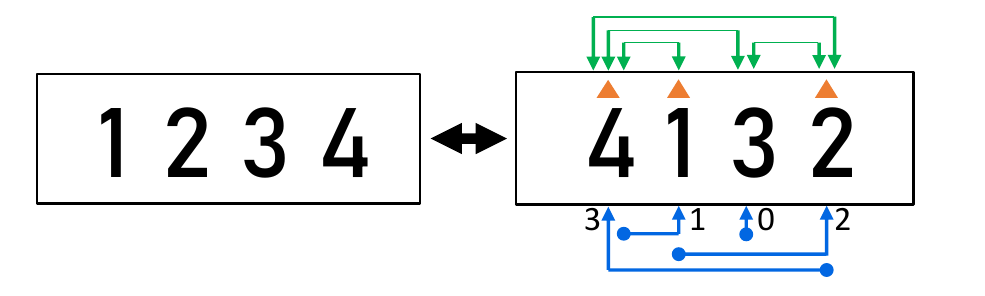}
    \caption{Illustration of similarity metrics between two permutations. The number of discordant pairs \textit{(top of right-hand box, green)} is the Kendall distance, the squared element movement \textit{(bottom of right-hand box, blue)} is the Spearman's rank correlation, and the number of changing elements \textit{(triangles, orange)} is the Hamming distance.}
    \label{fig:permutation}
\end{figure}

\paragraph{Permutation parameters}
Permutation parameters are used to describe the reordering of a sequence of elements. In compiler applications this most commonly appears as the reordering of a set of loops~\cite{haj2020autophase}. Consider for example a kernel with four nested loops $(l_1, l_2, l_3, l_4)$ which can be performed in any order. This ordering can be represented by a single permutation variable $\bm{\pi}$, which is a vector whose element $i$, given by $\pi_i=j$, describes the index $j$ of loop $l_i$ in the new order. For example, the permutation $\bm{\pi} = [2,4,3,1]$ corresponds to the following loop reordering:
\begin{minipage}[c]{\linewidth}
\begin{minipage}[c]{0.4\linewidth}
\begin{lstlisting}
for (l1 ... )
    for (l2 ... )
        for (l3 ... )
            for (l4 ... )
                ...
\end{lstlisting}
\end{minipage}
\hfill
\begin{minipage}[c]{0.15\linewidth}
\centering
$ \rightarrow $
\end{minipage}
\hfill
\begin{minipage}[c]{0.4\linewidth}
\begin{lstlisting}
for (l4 ... )
    for (l1 ... )
        for (l3 ... )
            for (l2 ... )
                ...
\end{lstlisting}
\end{minipage}
\end{minipage}
Prior black-box optimization literature, autotuning, and DSE frameworks lack the capability to effectively handle this variable type,  with the notable exception of OpenTuner~\cite{opentuner14}. 
In BO frameworks employing GPs, it is important to accurately estimate how different permutations relate to each other. In other words, for the nested loop reordering example above, the framework needs to determine if the two different loop orderings are likely to yield a similar performance. One naive way of handling permutation variables is to treat them as categorical variables, e.g., to consider one nested loop ordering to be equally similar to every other loop ordering (with the exception of itself). This, however, ignores the underlying structure that can be used to define a more refined similarity measure. We instead present three different semimetrics for permutations: the \textit{Kendall distance}, \textit{Spearman's rank correlation}, and the \textit{Hamming distance}. While the semimetrics are not strictly distance metrics, Lomelí \etal~\cite{lomeli2019antithetic} show that they can be used to form a valid GP kernel. These three semimetrics are illustrated in Fig.~\ref{fig:permutation} on a set of four elements, where the two boxes represent two permutations $\pi = [1,2,3,4]$ and $\pi'=[2,4,3,1]$. 

The first semimetric is the Kendall distance, $d_{k}(\pi, \pi') = \sum_{i=1}^{m}\sum_{j=1}^m |\mathbbm{1}_{\pi_i'>\pi_j} - \mathbbm{1}_{\pi_i>\pi_j'}|$. The Kendall distance represents the number of discordant pairs, i.e., the elements that have swapped order between the two permutations. Each discordant pair is represented by a green, interconnected double arrow. The second semimetric is the Spearman's rank correlation, $d_{s}(\pi, \pi') = \sum_{i=1}^m(\pi_i - \pi_i')^2$, which is the sum of squared movements of the elements between two permutations. It is illustrated with blue arrows in Fig.~\ref{fig:permutation}, where the dots represent the starting points and the arrows the final position.
For example, the number two starts in the second position in $\pi$ (left) and moves to the fourth position in $\pi'$ (right), meaning that it has travelled a distance of two. The Spearman's rank correlation then sums the squared displacement of all elements. Note that the square substantially emphasizes large rank changes. Lastly, the Hamming distance, $d_{h}(\pi, \pi') = \sum_{i=1}^m\mathbbm{1}_{\pi_i\neq\pi'_i}$, is the number of elements in $\pi$ that are no longer at their original position in $\pi'$ -- represented with orange triangles in the figure.

For a given permutation set, the choice of semimetric depends on how those permutation parameters are expected to impact the performance metric. Intuitively, Kendall distance focuses more on parameter order, whereas Spearman's rank correlation emphasizes large movements of individual elements. The Hamming distance only considers the number of elements changed and ignores where they moved to. As an example, consider the two loop orders
\begin{minipage}[c]{\linewidth}
\begin{minipage}[c]{0.4\linewidth}
\begin{lstlisting}
for (l2 ... )
    for (l3 ... )
        for (l1 ... )
            for (l4 ... )
                ...
\end{lstlisting}
\end{minipage}
\begin{minipage}[c]{0.1\linewidth}
\centering
and
\end{minipage}
\begin{minipage}[c]{0.4\linewidth}
\begin{lstlisting}
for (l4 ... )
    for (l3 ... )
        for (l1 ... )
            for (l2 ... ).
                ...
\end{lstlisting}
\end{minipage}
\end{minipage}
They have a high Spearman's rank correlation due to the large movement of the first and last element and relatively smaller Kendall and Hamming distances, which is intuitive given the compiler transformation that this represents.
This is backed by our ablation analysis in Sec.~\ref{sec:results}, where we observe that using Spearman's rank correlation outperforms the other alternatives. 
By consequence, we use Spearman's rank correlation as a default setting for permutation variables in BaCO.

\subsection{Parameter constraints}
\label{sec:constraints}

For an autotuning framework to be truly competitive in the complex world of modern autoscheduling, it is essential to effectively handle constraints in the parameter search space. Constraints can be divided into known constraints, which are known prior to optimization, and hidden constraints, which are only discovered during optimization. BaCO is designed to support both these constraint types.

\paragraph{Known constraints}
\label{sec:constrained}

In autotuning applications, users often possess expert knowledge regarding parameter configurations that lead to inefficient or even infeasible schedules. Incorporating this knowledge into the autotuning framework leads to significant performance improvements. The improvement becomes even greater when the feasible set makes up a small fraction of all possible configurations, i.e., when the search space is sparse. BaCO handles known constraints during the acquisition function optimization, and proposes only feasible configurations. As such, the surrogate model trains exclusively on feasible points.
 
BaCO uses a \textit{Chain of Trees} (CoT) data structure to deal with sparse search spaces, which was first presented by Rasch \etal~\cite{RaschSSG21}. The CoT computes all of the feasible configurations a-priori and stores them as a collection (or ``chain'') of trees. Each tree corresponds to a group of co-dependent parameters, and parameters in different trees are independent of one another. For each tree, each level of the tree corresponds to a single parameter and each node in that level corresponds to a possible value for that parameter. Each path from the root to a leaf then represents a partial configuration, and the tree is built so that only feasible configurations are included. Consider for example the following search space:
{
\newcommand\scalemath[2]{\scalebox{#1}{\mbox{\ensuremath{\displaystyle #2}}}}
$$
\scalemath{0.85}{
    p_1 \in \{2,4\},~p_2 \in \{2,4\},~p_3 \in \{1,4\},~p_4 \in \{1,2,4\},~ p_5 \in \{2,4,8\}
}
$$
$$
\scalemath{0.85}{
    p_1 \ge p_2,~ p_4 \ge p_3,~ p_5 \ge 2p_4
}
$$
}

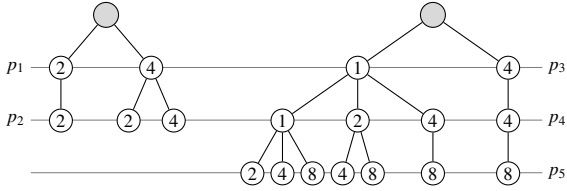
\begin{figure}
    \centering
\begin{tikzpicture}
 [level distance=2em, every node/.style={shape=circle,draw,align=center}, every node/.append style={font=\scriptsize}]
 \tikzstyle{level 0}=[every node/.style={shape=circle,draw,align=center}]
 \tikzstyle{level 1}=[sibling distance=12mm,inner sep=1pt] 
 \tikzstyle{level 2}=[sibling distance=6mm,inner sep=1pt] 
\node [fill=gray!30] (S){}
    child {node [fill=white!30] {2}
        child {node [fill=white!30] {2}}
    }
    child {node [fill=white!30] {4}
        child {node [fill=white!30] {2}}
        child {node [fill=white!30] {4}}
    };
[level distance=2em, every node/.style={shape=circle,draw,align=center,inner sep=1pt}, every node/.append style={font=\scriptsize}]
 \tikzstyle{level 1}=[sibling distance=20mm,inner sep=1pt] 
 \tikzstyle{level 2}=[sibling distance=10mm,inner sep=1pt] 
 \tikzstyle{level 3}=[sibling distance=4mm,inner sep=1pt] 
\node [right=4cm of S, fill=gray!30]{}
    child {node [fill=white!30] {1}
        child {node [fill=white!30] {1}
            child {node [fill=white!30] {2}}
            child {node [fill=white!30] {4}}
            child {node [fill=white!30] {8}}
        }
        child {node [fill=white!30] {2}
            child {node [fill=white!30] {4}}
            child {node [fill=white!30] {8}}
        }
        child {node [fill=white!30] {4}
            child {node [fill=white!30] {8}}
        }
    }
    child {node [fill=white!30] {4}
        child {node [fill=white!30] {4}
            child {node [fill=white!30] {8}}
        }
    };
\draw (-1.2, -.7) node[draw = none]{$p_1$};
\draw (-1.2, -1.4) node[draw = none]{$p_2$};
\draw (6, -.7) node[draw = none]{$p_3$};
\draw (6, -1.4) node[draw = none]{$p_4$};
\draw (6, -2.1) node[draw = none]{$p_5$};

\begin{scope}[on background layer]
    \draw [gray](-1,-.7) -- (5.8,-.7);
    \draw [gray](-1,-1.4) -- (5.8,-1.4);
    \draw [gray](-1,-2.1) -- (5.8,-2.1);
\end{scope}
\end{tikzpicture}
    \caption{Chain-of-Trees with 5 parameters, $p_1$ to $p_5$.}
    \label{fig:CoT}
\end{figure}
In this example, there are five input parameters and three constraints. Parameters $p_1$ and $p_2$ as well as parameters $p_3$, $p_4$, and $p_5$ are co-dependent. We represent them with the CoT shown in Fig.~\ref{fig:CoT}. As the parameters in different trees are independent, any combination of partial feasible configurations from the different trees yields a feasible configuration. For example, the leftmost path in the left tree combined with the rightmost path in the right tree yields the feasible configuration
\vspace{-.75em}
$$(p_1, p_2, p_3, p_4, p_5) = (2, 2, 4, 4, 8).\vspace{-.5em}$$

The use of the Chain-of-trees in BaCO is threefold. First random sampling can be made directly from the CoT. Secondly, instead of evaluating the constraints explicitly, it is significantly faster to instead check whether the configuration belongs to the CoT. Thirdly, it allows working with highly sparse search spaces which are common in the autotuning domain. In such sparse spaces, operating directly on the original domain becomes infeasible.\footnote{In the results section, we have used domain expert knowledge to manually transform the search spaces to limit the sparsity. This allows a more interesting comparison with previous methods.}

Working with constrained spaces is inherently biased since the optimization method presented in Rasch~\etal~\cite{rasch2017atf} prioritizes different configurations depending on the structure of the CoT. Their approach is equivalent to random sampling a configuration from the CoT by starting at each root node and then iteratively choosing a child with uniform probability, which is biased towards configurations in less dense parts of the tree. This bias is furthermore dependent on the order in which the parameters appear in the tree, which is an undesirable feature. We propose an alternative approach where we instead sample uniformly from the leaves of the trees, which is bias-free, meaning that the random sampling will be performed uniformly on the all the configurations. 
We study the impact of the bias in Sec.~\ref{sec:evaluation}.

Another source of sparsity comes from how parameters are defined prior to optimization. In autotuning and DSE applications, due to the nature of how computers store information, parameters are often constrained to take exponential values. Treating such parameters as integers leads to sparsity in the search spaces. BaCO instead applies the logarithmic transformation to such parameters. This transformation makes the search space significantly denser and yields more natural distances for the GP. These qualities improve performance, as we shall see in~Sec.~\ref{sec:evaluation}.

\paragraph{Hidden constraints}
Requiring the complete feasible domain definition from the user would severely limit the autotuner's usability. Some constraints are either too complicated to describe analytically or unknown a-priori. Instead, BaCO supports the concept of hidden constraints, learned automatically during optimization. BaCO uses a Random Forest model to predict the probability of feasibility for each configuration. It then extends the EI acquisition function presented in Sec.~\ref{sec:acquisition}, by multiplying the EI with the probability of feasibility \cite{nardi2019}. This should be compared with the naive approach of assigning high objective values to infeasible configurations, which suffers from difficulties with setting an accurate penalty. Such penalty terms are further often detrimental for the statistical model.

However, the practical interaction between the acquisition function based on the GP model and the RF feasibility predictor is complex. There is a constant trade-off between the feasibility model wanting to select feasible points and the value predictor that seeks to explore the unexplored infeasible regions. If the surrogate model becomes excessively confident within the feasible region, this balance tends to be skewed. In which case the selection fails to reliably find feasible points. As a practical solution, we consider a minimum feasible limit $\varepsilon_{f}$ and only consider configurations with probability of feasibility greater than $\varepsilon_f$ for evaluation. By randomly sampling a new $\varepsilon_f$ each iteration, with $p(\varepsilon_f=0) > 0$, we asymptotically guarantee to not cut away any solutions by doing this.

\section{Evaluation}

\label{sec:evaluation}
We validate the efficiency, effectiveness, and generalizability of BaCO. We first introduce the reference autotuning methods that we use as a baseline to evaluate the performance of BaCO, followed by the benchmarks from the three real-world frameworks presented in Sec.~\ref{sec:complexity}. We then show the performance results. 
For lack of space we show the extensive empirical results on all the frameworks and benchmarks in Appendix~\ref{app:additional_results}. All experiments are run for 30 repetitions.
We also show a wall-clock time analysis of all the autotuners used in Appendix~\ref{sec:wallclock}.

We answer the following research questions (RQ):

\textit{RQ1) Does \ourmethod achieve high performance with a limited autotuning budget?}
The evaluation in Fig.~\ref{fig:frameworks_geomean_performance} shows that, with a tiny budget of 20--40 evaluations, depending on the complexity of the benchmark, \ourmethod achieves
1.35$\times$--1.55$\times$
better performance than the state-of-the-art baselines.
Furthermore, BaCO consistently achieves expert-level performance with a small budget of 40--80 evaluations, where the baselines struggle to achieve expert-level performance even with a much larger budget.
This demonstrates the advantage of BaCO to deliver high performance for a small budget, even for complex search spaces.

BaCO delivers expert-level performance on average 2.87$\times$--3.87$\times$ faster than the 
state-of-the-art autotuning frameworks. Fig.~\ref{fig:eval} highlights the quicker performance evolution of BaCO for representative benchmarks, 
and a more detailed breakdown is presented in Table~\ref{tab:faster_performance} in the appendix. These results show that BaCO delivers performance much quicker than the baselines.

\textit{RQ2) Does BaCO generalize across compiler frameworks and benchmarks?}
Our evaluation across three diverse real-world compiler frameworks shows consistently that BaCO significantly outperforms the baselines.
In fact, BaCO is the \emph{only} framework that outperforms the expert configuration on all benchmarks across compiler frameworks, as shown in Fig.~\ref{fig:eval}, with a more detailed breakdown in Table~\ref{tab:expert_level_achieved} in the appendix.
This observation suggests that the techniques discussed in the paper generalize well across compiler frameworks and benchmarks.

\textit{RQ3) What is the performance benefit of customizing Bayesian optimization for compiler autotuning?}
Our system demonstrates advantages over prior work BO-based autotuning frameworks not specifically customized for compiler domains (Ytopt in Fig.~\ref{fig:abl}). As explained in Sec.~\ref{sec:framework} and Sec.~\ref{sec:search_spaces}, 
these improvements validate our design choices and suggests that there are significant performance benefits to be gained by customizing the BO framework for this particular domain.

\textit{RQ4) What are the findings of autotuning our distinct real-world compilers using \ourmethod?} 
This question attempts to provide insight on why \ourmethod outperforms baselines for our real-world compiler benchmarks. We identify three main areas: exploration of new configurations, testing schedules that did not previously exist in prior work, and better handling of both known and unknown constraints for complex real-world applications. Evaluations for this question comes from Fig.~\ref{fig:frameworks_geomean_performance}, Fig.~\ref{fig:line}, and Fig.~\ref{fig:eval:complex} in the Appendix.

\subsection{Baseline Methods}
\label{sec:baselines}
To contextualize the performance of BaCO, we evaluate it alongside two state-of-the-art autotuning frameworks and two random sampling approaches.

\textbf{ATF with OpenTuner}
The Auto-Tuning Framework (ATF)~\cite{RaschSSG21}  extends the popular  OpenTuner~\cite{opentuner14} to handle known constraints. 
We chose OpenTuner as a baseline since it is one of the leading frameworks for autotuning.

\textbf{Ytopt} 
Ytopt~\cite{wu2021autotuning} is an autotuning framework using BO and is part of the PROTEAS-TUNE project~\cite{proteastune}. It supports both Random Forests and Gaussian processes. We run it here with Random Forests as the GP implementation does not support constraints. When infeasible solutions are found due to hidden constraints, they are added to the data set with a high objective value. We compare against Ytopt since it is one of the only frameworks that supports either constraints or GP.
When addressing RQ3, we run Ytopt with GPs.

\textbf{Uniform and CoT random sampling}
These are uniform random sampling methods. The CoT random sampling is a method that randomly samples at uniform directly from the CoT. 
This baseline allows us to study the impact of the bias introduced by the known constraints, as explained in Sec.~\ref{sec:constrained}.

\textbf{Default and expert configurations}
For reference, we show the performance of two baseline configurations: The \textit{default} configuration, and an \textit{expert} configuration, carefully handcrafted by domain experts in the respective programming languages. It is unlikely that a developer would exceed the expert performance baseline, which makes it a suitable data point for our empirical analysis. 
The HPVM2FPGA benchmarks are automatically generated by the autoscheduler and do not provide any expert configurations, in which case we only report the default. 
The expert configurations are taken from prior publications: TACO~\cite{senanayake2020scheduling} and RISE~\&~ELEVATE~\cite{hagedorn2020achieving, koehler2021towards, steuwer2015generating, stoltzfus2019}. The original authors used manual or semi-automated methods to determine well-performing configurations based on their experience, hardware characteristics, or data properties.
The authors had the incentive to produce the best configurations but, presumably due to time constraints, they may have occasionally missed better-performing configurations.

\begin{table}
    \centering
    \footnotesize
    \addtolength{\tabcolsep}{-0.4em}
    \begin{tabular}{r|cccccc}
        \textbf{Benchmark} & \multicolumn{6}{l}{\textbf{Dim}\hspace{.6cm}  \textbf{Constr.}\hspace{1.35cm} \textbf{Feasible}\hspace{.5cm}  \textbf{Full}} \\
         & \multicolumn{6}{l}{\strut\hspace{.4cm}\textbf{Params}\hspace{.6cm} \textbf{Space size} \hspace{1.45cm} \textbf{Budget}} \\
        \hline\multicolumn{6}{c}{\textbf{TACO}} \\
        \hline SpMV & 7 & O/C/P &  & $1.5 \times 10^{7}$ & $3.0 \times 10^{6}$ & 70\\
        SpMM & 6 & O/C/P & K & $5.2\times10^{11}$ & $4.7\times10^4$ & 60\\ 
        SDDMM & 6 & O/C/P & K & $5.2\times10^{11}$ & $7.8\times10^4$ & 60\\ 
        TTV & 7 & O/C/P & K/H & $1.5 \times 10^{7}$ & $6.0 \times 10^{6}$ & 70\\
        MTTKRP & 6 & O/C/P & K & $1.5 \times 10^{6}$ & $6.8 \times 10^{5}$ & 60\\
        \hline\multicolumn{6}{c}{\textbf{RISE\,\&\,ELEVATE}} \\
        \hline 
        MM\_CPU & 5 & O/P & K/H & $1.0 \times 10^{7}$ & $2.9\times10^4$ & 100\\ 
        MM\_GPU & 10 & O & K/H & $1.1 \times 10^{11}$ & $1.5 \times 10^8$  & 120\\ 
        Asum\_GPU & 5 & O & K & $1.2 \times 10^{6}$ & $6.3\times10^4$  & 60 \\ 
        Scal\_GPU & 7 & O & K/H & $3.9 \times 10^{7}$ & $4.2 \times 10^6$  & 60\\
        K-means\_GPU & 4 & O & K/H & $1.4 \times 10^{4}$ & $3.6 \times 10^3$  & 60\\
        Harris\_GPU & 7 & O & K & $7.7 \times 10^{9}$ & $1.0 \times 10^{7}$ & 100\\        
        Stencil\_GPU & 4 & O & K & $1.4 \times 10^{4}$ & $3.6 \times 10^{3}$ & 60 \\
        \hline\multicolumn{6}{c}{\textbf{HPVM2FPGA}} \\
        \hline BFS & 4 & I/C & H & $256$ & $256$ & 20\\
        Audio & 15 & I/C & H & $8.4\times 10^5$ & $8.4\times 10^5$ & 60\\
        PreEuler & 7 & I/C & H & $1.5\times10^4$ & $1.5\times10^4$ & 60\\\hline
    \end{tabular}
    \caption{We evaluate \ourmethod on 15 important kernels from domains like machine learning, statistics, and signal processing. The benchmarks expose search spaces with varying number of parameters (\textit{Dim}). They cover all parameter types considered (\textit{Params}): real (R), integer (I), ordinal (O), categorical (C), and permutation (P).
    \textit{Constr.} describes the type of constraints used by the benchmark: known (K) and hidden (H) constraints. \textit{Space size} describes the number of possible configurations in the dense search space with \textit{Feasible} denoting all valid configurations with respect to the known constraints. \textit{Full Budget} is the total number of evaluations we allow for autotuning the kernel. }
    \label{tab:benchmarks}
\end{table}

\subsection{Benchmarks}
BaCO is evaluated over 15 kernels from linear algebra, machine learning, image processing, statistics, and signal processing.
We integrate BaCO in the three real-world frameworks presented in Sec.~\ref{sec:complexity}. 
The benchmarks have been chosen based on prior work by the authors of the three frameworks. Furthermore, most of these benchmarks have an expert optimized code which allows for a fair comparison. The search space size ranges from tens of thousands to billions of configurations, as described in Table~\ref{tab:benchmarks}, which is beyond the scope of exhaustive search.

To define the evaluation budget for each benchmark, we first establish a full budget for each benchmark, as shown in the last column of Table~\ref{tab:benchmarks}. 
The full budget is defined using a rule of thumb of around $5$ to $6$ minutes. This max compilation time is commonly adopted in large companies, such as Google. We then define \emph{tiny} and \emph{small} budgets as 1/3 and 2/3 of the \emph{full} budget.

\textbf{TACO benchmarks}
We benchmark 5 tensor algebra expressions, commonly used in machine learning and tensor factorization~\cite{sparsekernels2020}. Namely, they comprise of sparse matrix-vector multiply (SpMV) $a_i = \sum_k B_{ij}c_k$, sparse matrix multiply (SpMM) $A_{ij} = \sum_k B_{ik}C_{kj}$, sampled dense-dense matrix multiply (SDDMM) $A_{ij} = \sum_k B_{ij}C_{ik}D_{jk}$, tensor times vector (TTV) $A_{ij} = B_{ijk}c_k$, and fourth-order matricized tensor times Khatri-Rao product (MTTKRP) $A_{ij} = B_{iklm}*C_{kj}*D_{lj}*E_{mj}$.
Each tensor expression is given a scheduling template that exposes tiling parameters (split and unrolling factors) and permutation parameters (loop reorderings). BaCO searches for the set of parameters, and therefore the schedule, that yields the best performance. The characteristics of these parameters and the search space is described in Table~\ref{tab:benchmarks}. We use tensors from a wide variety of real-world applications ranging from power networks and circuits to fluid dynamics and social networks. We run matrix expressions on a subset of SuiteSparse matrices~\cite{davis2011university, kolodziej2019suitesparse} and synthetic uniform random tensors, and we run higher-order expressions on the Facebook Activities tensor~\cite{viswanath2009evolution}, a subset of the FROSTT tensor collection~\cite{frosttdataset}, and synthetic tensors as well (see Table~\ref{tab:dataset}). The selected tensors vary widely across tensor properties including number of nonzeroes, dense dimension size, and irregular nonzero pattern.

The TACO benchmarks were run on nodes with two Intel Xeon Gold 6130 processors locked at 2100Ghz, using all 32 cores and 96GB of RAM.

\begin{table}[tb]
    \centering
    \scriptsize
    \setlength{\tabcolsep}{2.8pt}
    \begin{tabular}{r|ccc}
        \textbf{Matrix} & \textbf{Dimension} & \textbf{Nonzeroes} & \textbf{Dataset} \\
        \hline ACTIVSg10K & $20,000\times20,000$ & $135,888$ & SS\\
        email-Enron& $36,692\times36,692$ & $367,662$ & SS\\
        Goodwin\_040 & $17,922\times17,922$ & $561,677$ & SS\\
        scircuit & $170,998\times170,998$ & $958,936$ & SS \\
        filter3D & $106,437\times106,437$ & $2,707,179$ & SS\\
        laminar\_duct3D & $67,173\times67,173$ & $3,788,857$ & SS\\
        cage12 & $130,228 \times 130,228$ & $2,032,536$ & SS\\
        smt & $25,710 \times 25,710 $ & $3,749,582$ & SS\\
        random2 & $10,000 \times 10,000$ & $5,000,000$ & Rand\\
        random1 & $1000 \times 500 \times 100$ & $5,000,000$ & Rand\\
        facebook & $1,504 \times 42,390 \times 39,986$ & $737,934$ & FB\\
        uber & $183 \times 24 \times 1,140 \times 1,717$ & $3,309,490$ & FT\\
        nips & $2,482 \times 2,482 \times 14,036 \times 17$ & $3,101,609$ & FT\\
        chicago & $6,186 \times 24 \times 77 \times 32$ & $5,330,673$ & FT\\
        uber3 & $183 \times 1,140 \times 1,717$ & $1,117,629$ & FT*\\
    \end{tabular}
    \caption{Tensors used in our TACO benchmarks from the SuiteSparse matrix collection (SS)~\cite{davis2011university, kolodziej2019suitesparse}, Facebook Activities graph (FB)~\cite{viswanath2009evolution}, FROSTT (FT) collection~\cite{frosttdataset} or synthetically generated (Rand). \\ 
    *We modify a FROSTT 4-tensor to a 3-tensor by dropping one dimension since the next largest FROSTT tensor has more than  $75\times$ the number of nonzeros.}
    \label{tab:dataset}
\end{table}

\textbf{RISE\,\&\,ELEVATE benchmarks}
We use six benchmarks covering multiple domains, optimizations, and hardware devices. This results in benchmarks requiring various autotuning features, as described in Table~\ref{tab:benchmarks}.

The CPU Matrix Multiplication (MM\_CPU) benchmark from \cite{hagedorn2020achieving} is run on a CPU and applies tiling, vectorization, and loop-permutation optimizations. 
The remaining benchmarks are run on a GPU and apply GPU-specific optimizations, including the OpenCL-specific work-group configuration, memory hierarchies, and coalescing. 
The MM\_GPU and K-means\_GPU dense linear algebra benchmarks are inspired by implementations used in \cite{steuwer2017lift}. 
The linear algebra algorithms Asum\_GPU and Scal\_GPU are from \cite{steuwer2015generating}, the stencil from \cite{stoltzfus2019}.
The remaining image processing algorithm Harris\_GPU is a corner detector described in \cite{koehler2021towards}. 

The RISE\,\&\,ELEVATE evaluation was executed on 8 cores of an Intel Xeon E5-2650 v3 @2.30Ghz processor accompanied by 32 GB of RAM.
For the GPU benchmarks, we used a NVIDIA K80 GPU.

\textbf{HPVM2FGA benchmarks}
We use the benchmarks presented in~\cite{ejjeh2022hpvm2fpga}: (1) Breadth First Search (BFS), and (2) the computational fluid dynamics algorithm of euler with pre-computed fluxes (PreEuler), are taken from the Rodinia Benchmark suite~\cite{che2009rodinia}, and (3) 3D Spacial Audio Encoder (Audio) from the ILLIXR testbed~\cite{huzaifa2021illixr}. 
The benchmarks represent diverse workloads from different domains with varying parameter space sizes, ranging from 4 parameters for BFS, to 15 for Audio. 

We ran these benchmarks through HPVM2FPGA's optimizer, reporting the estimated execution time targeting an Intel Arria 10 GX FPGA in our evaluation results.

\begin{figure}
    \centering
    \includegraphics[width=0.5\textwidth]{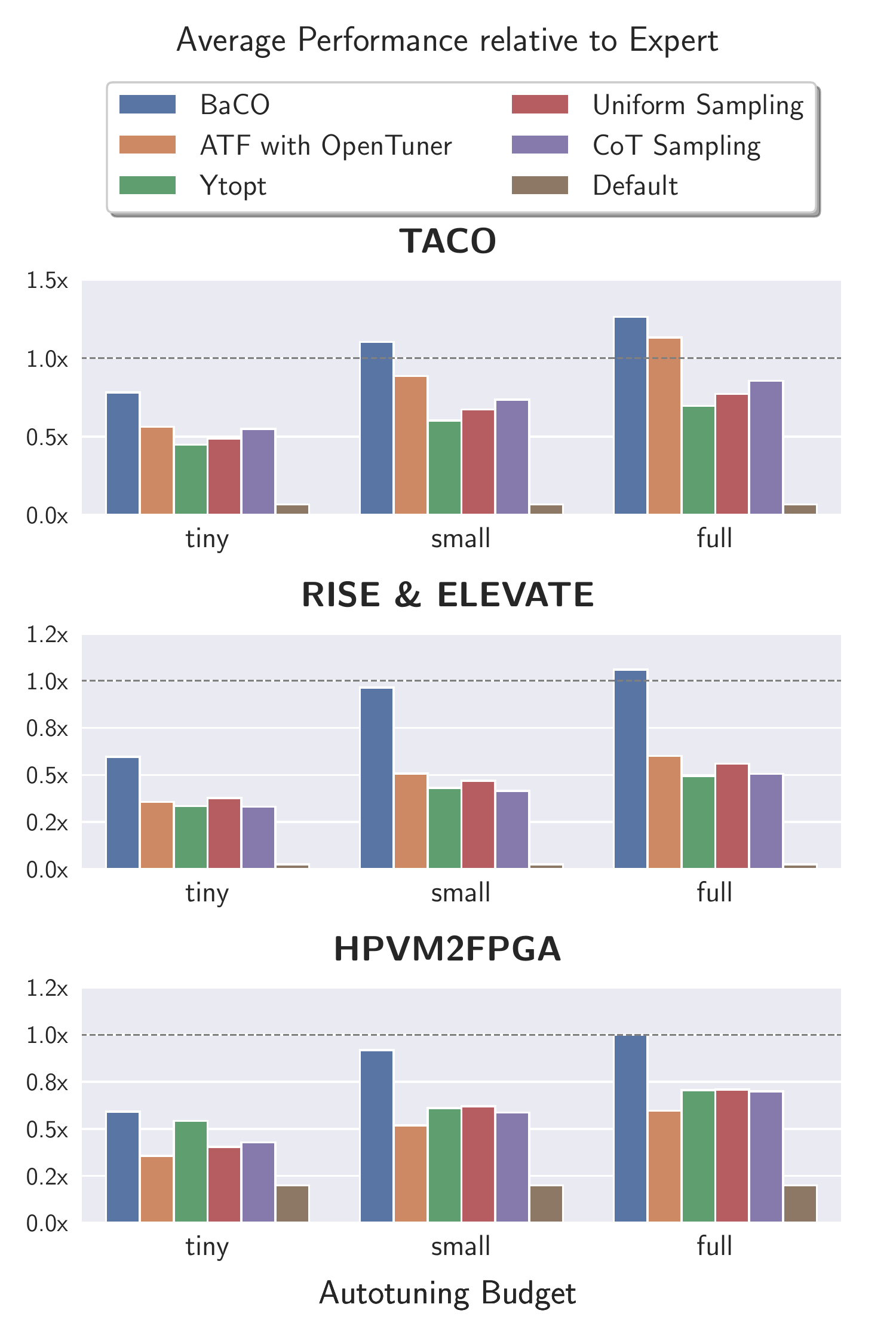}    
    \caption{Average performance relative to expert with a tiny budget (1/3 of the full budget), a small budget (2/3 of the full budget) and the full budget.
    BaCO delivers the highest performance for each budget and achieves expert-level performance consistently with the small budget.
    }
    \label{fig:frameworks_geomean_performance}
\end{figure}

\subsection{Results}
\label{sec:results}

\begin{figure*}[t]
    \centering
    \includegraphics[width =\textwidth]{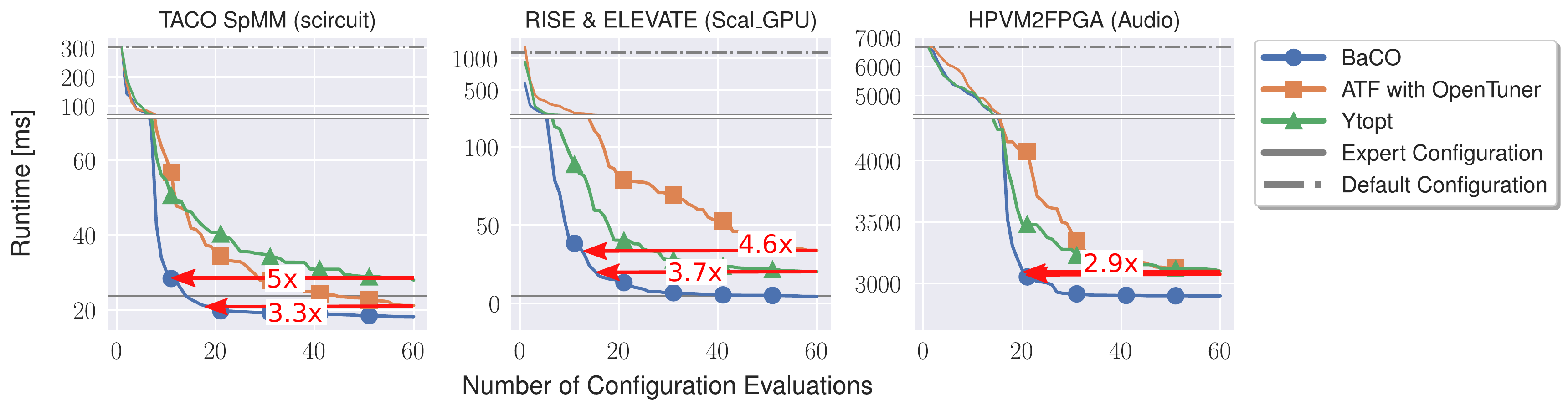}
    \caption{Evolution of average best runtime for one kernel from each framework. The figure is split vertically into two different scales. BaCO reaches the final performance of the state-of-the-art methods using as little as 30\% of the function evaluations of the other methods.}
    \label{fig:eval}
    \label{fig:quicker_performance}
\end{figure*}

\textbf{RQ1) Does \ourmethod achieve high performance with a limited autotuning budget?}
Fig.~\ref{fig:frameworks_geomean_performance} shows the average performance of BaCO and the baselines for three different levels of autotuning budget for all our benchmarks. 
The tiny budget is only 20-40 evaluations.\footnote{Besides the BFS benchmark, for which it is only 6 evaluations}
BaCO clearly outperforms the baselines, delivering on average better performance than all baselines for 19 out the 24 benchmarks.
For TACO, the tiny budget is even sufficient to deliver expert-level performance.
With the small budget, BaCO delivers expert performance for all three compiler frameworks.
The baselines struggle to deliver good performance even with the full budget, particularly for the challenging spaces in the RISE benchmarks. Tables~\ref{tab:relative_perf_33}, \ref{tab:relative_perf_66}, and \ref{tab:relative_perf_100} in the appendix show the detailed performance results for each individual benchmark and autotuning framework.

BaCO also achieves performance faster, i.e., it reaches the final performance of the other baselines using fewer configuration evaluations. Fig.~\ref{fig:quicker_performance} shows the performance evolution for three selected benchmarks, and that BaCO delivers performance with 2.9$\times$--5$\times$ fewer evaluations than ATF and Ytopt. On average, BaCO finds the best ATF and Ytopt configurations 2.87$\times$ and 3.82$\times$ faster, respectively. Our experiments confirm that these results generalize well across 
our benchmarks, however, due to space constraints these additional results are presented in Table~\ref{fig:quicker_performance} in the Appendix. 
 
\textbf{RQ2) Does BaCO generalize across compiler frameworks?} 
To see how the performance generalizes, we show in Fig.~\ref{fig:line} (and Fig.~\ref{fig:eval:complex} in the Appendix) how the mean of the best-found solution by each framework improves over time for each individual benchmark. In the figures, the average performance is plotted for each method and each benchmark. The goal is to achieve a lower value, which  identifies a better-performing configuration, and to find low values as far to the left as possible, which means using a low tuning budget. The performance of the default configuration and expert configuration is presented for reference when available. We further denote when each method reaches expert-level performance with a star, such that a shorter distance between the y-axis and the star is better. As finding improvement over the default configuration initially is easy, we split the plots into two regions with different scales. This helps focus on the interesting part closer to the expert-level performance. \ourmethod reliably yields high-level performance and overall provides the best schedule in 22 out of the 24 benchmarks. It is further frequently the only method to reach expert-level performance within the given budget.

\begin{figure*}[t!]
    \centering
    \centering
    \includegraphics[width=\textwidth]{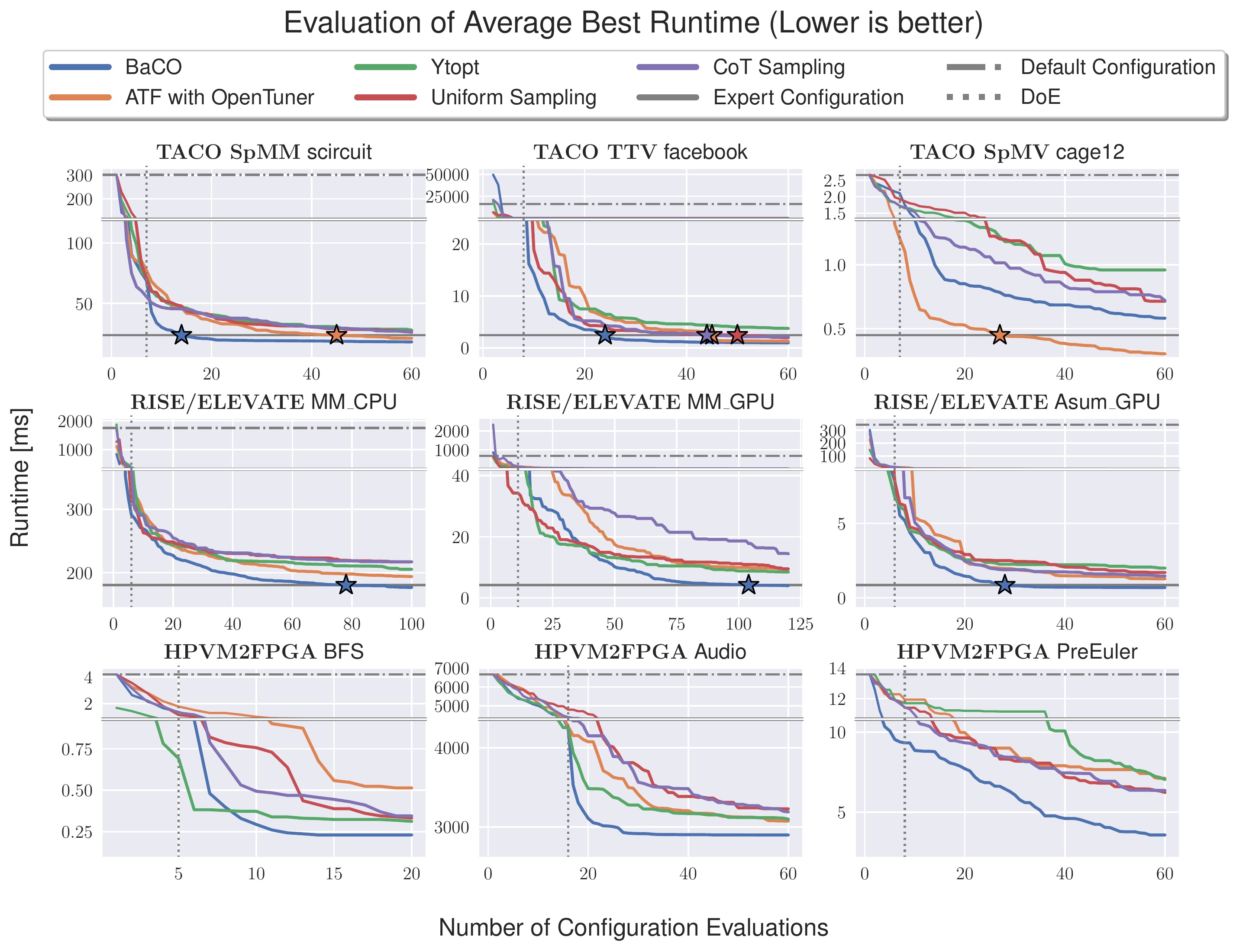}
    \caption{Evolution of average best runtime among evaluated configurations for selected benchmarks from Table~\ref{tab:benchmarks}. Fig.~\ref{fig:eval:complex} in the Appendix contains from Table~\ref{tab:benchmarks} not shown here. The figure is split vertically into two different scales, and we mark the iteration where each method beats the expert configuration with a star.}

    \label{fig:line}
\end{figure*}

\begin{figure}[]
    \centering
    \includegraphics[width=0.5\textwidth]{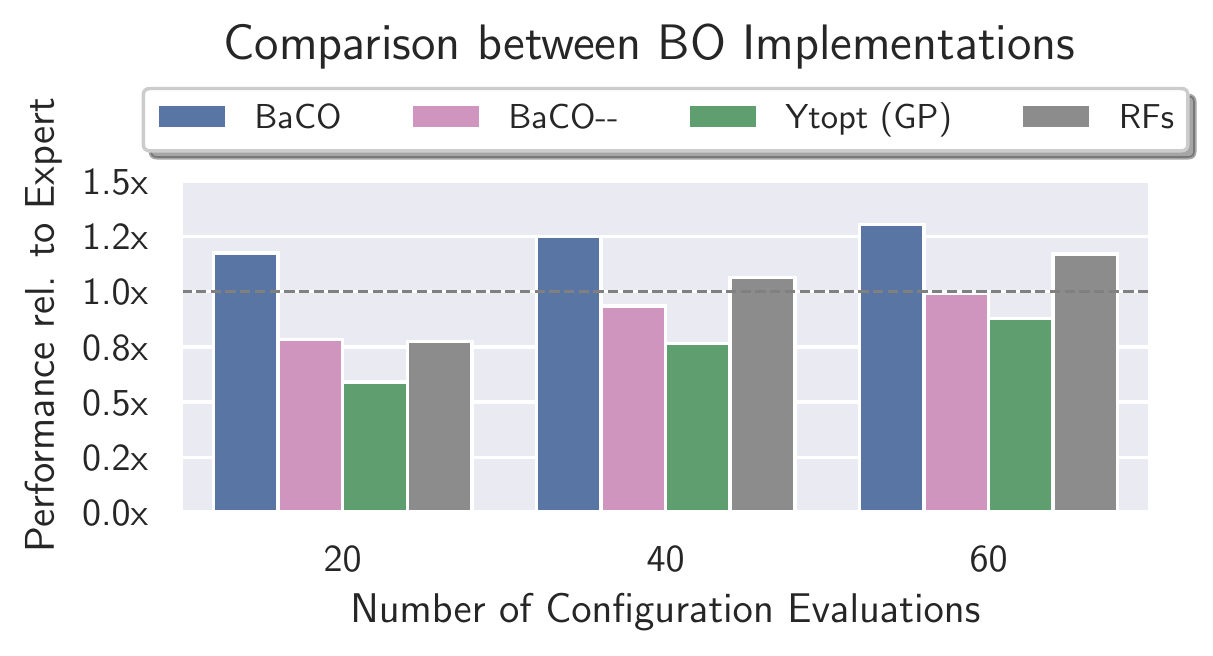}
    \caption{Geometric mean of the performance relative to the expert configuration for the TACO SpMM kernel applied to the \textit{filter3D}, \textit{email-Enron} and \textit{amazon0312} matrices after 20, 40 and 60 evaluations.}
    \label{fig:ablA}
    \label{fig:abl}
\end{figure}

\textbf{RQ3) What is the performance benefit of customizing Bayesian optimization for compiler autotuning?}
We study the \ourmethod design by running three matrices for SpMM with default settings and with a number of major features turned off. We use the SpMM benchmark as it is reasonably well-behaved and only has few constraints. 
The average speedup over expert is shown in Fig.~\ref{fig:ablA}. We denote the restricted version by \ourmethod-{}-. In more detail, \ourmethod-{}- is \ourmethod without variable transformations, model priors, and local search for the acquisition function. It further uses the naive distance for permutation variables that ignores their underlying structure and it does not use \ourmethod's more advanced fitting of the GP hyperparameters. We see that by doing those changes, \ourmethod takes about a 20\% performance loss.

Next, we compare it with the GP implementation of Ytopt. 
Ytopt uses none of the above mentioned features, but additionally has a less advanced GP and BO toolkit.
Ytopt only supports constraints for RF so this Ytopt configuration does not support constraints. However, for this benchmark we have manually pruned the search space for Ytopt, so that the only remaining constraint is a single less-than relationship between two variables. BO with GPs requires a lot of care to be efficient, which we see from the difference between Ytopt (GP) and \ourmethod-{}-.

Lastly, we show the difference between a well implemented GP and RF as surrogate model. Especially for smaller budgets, the GP model shows much stronger performance. This is relevant as there is currently a paradigm shift towards using GPs in discrete settings.

\begin{figure}
    \centering
    \includegraphics[width=0.5\textwidth]{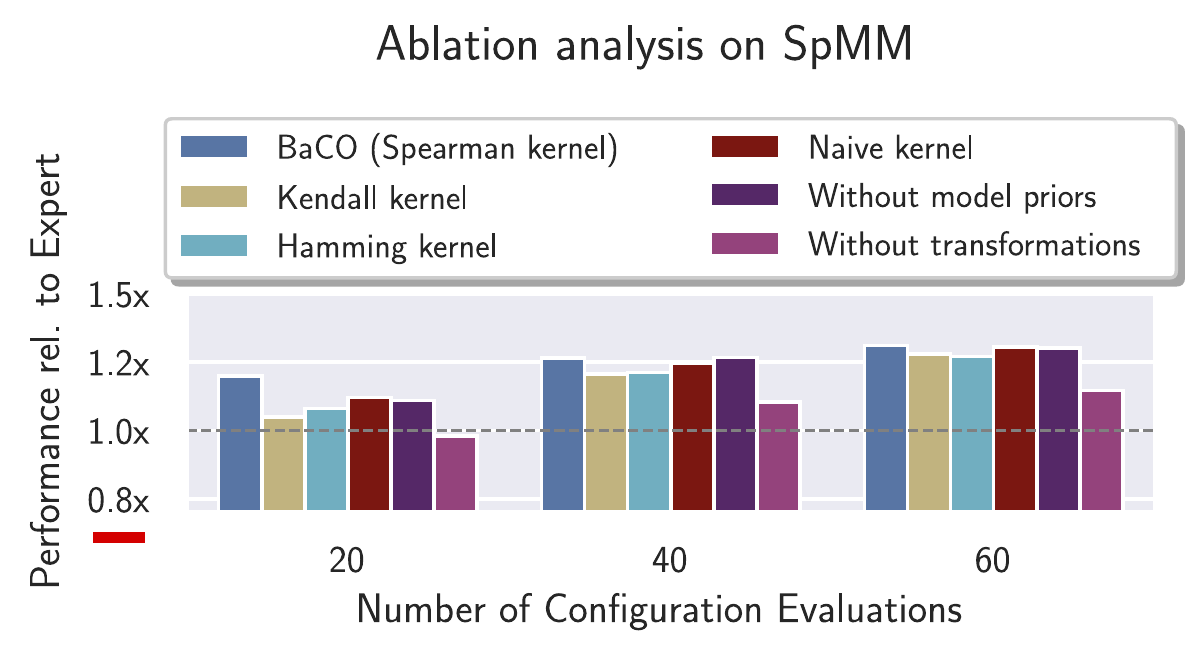}
    \caption{Geometric mean of the performance relative to the expert configuration for the SpMM kernel applied to the \textit{filter3D}, \textit{email-Enron} and \textit{amazon0312} matrices after 20, 40 and 60 evaluations. (Note the cut axis).}
    \label{fig:ablB}
\end{figure}

\textbf{Ablation analysis}
To further understand the impact of the different design choices in BaCO, we perform an ablation analysis in Fig.~\ref{fig:ablB}. The impact of the permutation kernel, variable transformations and model priors are studied in an ablation analysis. First, BaCO in default settings is presented, which is using Spearman's rank correlation for permutation variables. Then we study the impact of changing the permutation metric to Kendall distance, Hamming distance, as well as the naive approach of treating permutations as categorical variables (Sec.~\ref{sec:distance_metrics}). The Spearman metric yields the best performance, especially in early iterations.

Secondly, we study the impact of removing the logarithmic transformations of variables and output (Sec.~\ref{sec:constrained}), and the model priors (Sec.~\ref{sec:surrogate}). Removing the log transforms significantly deteriorates the performance at all evaluation counts. The lengthscale priors, however, have a larger impact early on, where they work to stabilize the procedure, and become less important when more data has been observed with which to fit the model. 

Overall, we see that the changes have a much larger impact early on. Except for transformations, ignoring any of the individual features fails to prevent good performance after more iterations. It is noteworthy that no individual design choice has a major impact, but together they make a large difference. 

\begin{figure}[]
    \centering
    \includegraphics[width=0.5\textwidth]{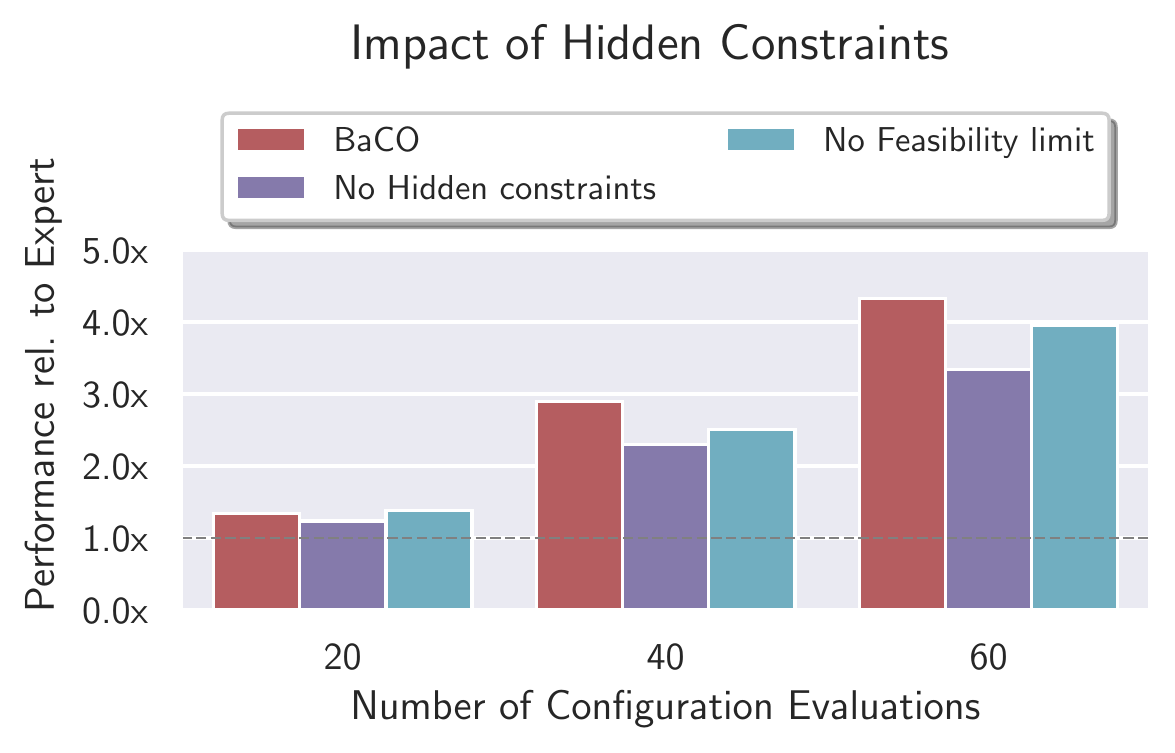}
    \caption{Geometric mean of the performance relative to the expert configuration for the MM\_GPU and Scal\_GPU kernels after 20, 40 and 60 evaluations.}
    \label{fig:hidden}
\end{figure}

\textbf{Hidden constraints}
Next, we study the impact of predicting feasibility with respect to hidden constraints on two benchmarks from the RISE/ELEVATE suite. In Figure \ref{fig:hidden}, we show the average improvement over expert after 20, 40, 60 iterations with and without the feasibility predictor. Additionally, we show the impact of the minimum feasibility limit presented in Sec.~4.2. It shows that the hidden constraints predictor has a significant positive impact, particularly after more iterations where it has had more samples to train on. But it also indicates that the introduced minimum feasibility limit (Sec.~\ref{sec:constraints}) is important to stabilize the interaction between the feasibility predictor and the surrogate model. 

\textbf{Chain-of-Trees}
Even after manual sparsity-reducing transformations, some of the search spaces remain highly sparse. When this is the case, CoTs greatly increases the efficiency of sampling from, and searching, the parameter domain. On the MM\_GPU search space for example, over ten runs, using the CoTs reduced the time spent on evaluating constraints in the local search by a factor $6\times$ and the random sampling by a factor of $80\times$. Overall, this resulted in that the time spent by the internal working of \ourmethod was reduced by $70\%$. For even more sparse search spaces, operating directly on the search space quickly becomes untenable.

\textbf{RQ4) What are the findings of autotuning our distinct real-world compilers using \ourmethod?} 

\textbf{Configuration insight} BaCO underperforms baselines in only 1 of our 24 (4\%) benchmarks across the 15 kernels. Opentuner is able to beat BaCO (as shown in Fig.~\ref{fig:line}) when running SpMV on cage12 (see Table~\ref{tab:dataset}). The SpMV benchmark is interesting as it has a good default setting, but ill-designed schedules can increase the run time by several orders of magnitude. After inspection of the configurations, it shows that ATF picks configurations similar to its previous configurations. This behavior in ATF exploits sampling around prior good configurations each evaluation, whereas BaCO's algorithm is more explorative in finding completely new configurations. Exploiting configurations works for simple kernels, like SpMV, but fails for real-world kernels with increased complexity and runtime variance. Exploitation sampling is likely to get stuck in a local minima, which is more likely to be globally bad for complex problems (as is the case with TTV on the random1 in Fig.~\ref{fig:eval:complex} in the Appendix). We do not augment BaCO in any specific manner to explore configurations, but the global nature of the BO paradigm emphasizes exploration over methods with local-exploration elements such as OpenTuner.

\textbf{Performance over expert} BaCO is able to achieve better than expert performance in some cases (see Fig.~\ref{fig:frameworks_geomean_performance}). Even experts in the domain, with insight of the underlying hardware architecture, may miss out on the optimal configuration and the best performance simply due to the amount of user-time needed to explore the vast search space of possible configurations. BaCO allows users to automate that search while potentially offering better performance than the expert could find. For example, BaCO is able to find over a $1.1\times$ speedup on average for TACO (see Fig.~\ref{fig:frameworks_geomean_performance}) since experts in~\cite{senanayake2020scheduling} only considered the default loop ordering (permutation) for all expressions. In addition, many of the configurations the autotuner discovers are hard to find by hand due to the concordant traversal of a compressed tensor data structure. Therefore, it is difficult for an expert to search the space of loop orderings and know which of them are infeasible.

\textbf{Constraints in real-world applications} Over half (8/15) of our benchmarks, and notably all of the HPVM2FPGA benchmarks, use hidden constraints. Additionally, all but one benchmark use known constraints, significantly reducing the feasible search space as shown by the Feasible column in Table~\ref{tab:benchmarks}. Predictor modeling of hidden constraints has a significant impact on performance (as discussed in RQ3), and this impact is apparent in our real-world compiler benchmarks.

\section{Related Work}

\textbf{Bayesian optimization for autotuning}
Several Bayesian optimization frameworks have been presented for autotuning~\cite{nardi2019, wu2021autotuning, willemsen2021, liu2021gptune}. One of the earlier frameworks was introduced by Nelson \etal~\cite{nelson2015generating}, who present \textit{SURF} that uses Random Forests models to optimize tensor constraction operations on GPUs. This work is later extended to become \textit{Ytopt} by Wu \etal \cite{wu2021autotuning}. The authors use the skopt Bayesian optimization framework to optimize LLVM Clang/Polly pragma configurations on the PolyBench benchmark suite. Ytopt further allows the usage of additional surrogate models such as Gaussian Processes and Boosted Trees.
Ytopt implements a method based on Bayesian optimization, which BaCO builds on. However, we show that the Bayesian optimization pipeline needs to be further customized to work well on autotuning domains, which is the scope of our work. 

The work by Sid-Lakhdar \etal~\cite{sid2019multitask} focuses on the meta- and multi-task learning aspect. It was extended by Liu \etal~\cite{liu2021gptune} into the \textit{GPTune} framework. The authors use \textit{linear coregionalization models} (LCMs), to model multiple similar problems simultaneously to increase efficiency. This was further extended by Zhu \etal~\cite{zhu2022gptuneband} to also handle multifidelity applications. 
While this is out of the scope of the current work, the use of meta-learning can be used in combination to BaCO to achieve greater efficiency. 
Willems \etal~\cite{willemsen2021} use Bayesian optimization to autotune GPU kernels using Gaussian Processes and known constraints on the search space. Another recent approach is \textit{Bliss} by Roy \etal~\cite{roy2021bliss}, that probabilistically chooses a combination of models and acquisition functions each new optimization iteration based on previous performance observations. 
\textit{Bliss'} approach is orthogonal to BaCO and it is possible that combining the methods further efficiency can be achieved. Recently, Dorier \etal~\cite{dorier2022hpc} present DeepHyper, a Bayesian optimization framework for HPC storage system autotuning, that focuses on transfer learning through the use of variational autoencoders. 

\textbf{Bayesian optimization for design space exploration}
Nardi \etal~\cite{nardi2019,koeplinger2018spatial} use Bayesian optimization with a RFs surrogate model to optimize FPGAs. They consider both multiobjective and hidden constraints.
Ejjeh \etal~\cite{ejjeh2022hpvm2fpga} use the same DSE framework to tune hardware-agnostic programs targeting FPGA backends.
\cite{souza2021bayesian} design a human-centric DSE approach, where expert priors accelerate the convergence of the autotuner. 
While this is not the focus of our work, a simple adaptation of the BaCO acquisition function can benefit the same user priors when available. 

There is substantial work in the literature about DSE techniques in HLS~\cite{bruel2017autotuning,ferretti2021graph,wu2021ironman,wang2021autosa,zhong2016linanalyzer}. However, most existing work focuses on using DSE for tuning HLS, rather than using it to select compiler optimizations~\cite{bruel2017autotuning,ferretti2021graph,zhong2016linanalyzer}. These works are not based on Bayesian optimization and we view them as complementary to our work. 

\textbf{The phase-ordering problem}
Autoscheduling tackles the task of  applying a number of transformations to optimize a kernel automatically. Typically, the scheduling language  parametrizes the application of those transformations by a bounded set of options which we refer to as parameters that are easier to optimize over. This parametrization approach is the one used by TACO and ELEVATE.
However, a different approach is to operate directly  on the space of transformations. Optimization over this unbounded tree-like space is commonly known as the \textit{phase-ordering problem}~\cite{kruse2020autotuning, koo2021customized, haj2020autophase, haj2020protuner}.

\section{Conclusions and Future Work}
We introduce the Bayesian Compiler Optimization framework (BaCO), a plug-and-play solution to autoscheduling tasks for modern scheduling languages targeting various hardware backends. BaCO is able to reach expert-level performance 2.7$\times$-10$\times$ faster than the state of the art autotuners.
The separation of concerns between policy and mechanism allows compiler users to delegate the complex and time-consuming task of scheduling to BaCO so that they can focus on their applications instead. 

While we show that BaCO can provide high-performing solutions in less than 100 seconds, this time is still too long for use in software development. The holy grail of autoscheduling is to be able to use an autotuner during the development, and ideally enable the user to run autotuning every time they compile their code. That way users can check both functional and non-functional properties on a regular basis during the various program lifecycle phases. Indeed, increasing the efficiency of the autotuner would enable a new level of autotuning-in-development-loop paradigm which is not accessible with the current state of autotuning technology.
\section*{Acknowledgements}
We would like to thank Jaeyeon Won for his help with evaluation and Michael O'Boyle and Tobias Grosser for their valuable feedback the a draft of this paper.
Luigi Nardi was supported in part by affiliate members and other supporters of the Stanford DAWN project — Ant Financial, Facebook, Google, Intel, Microsoft, NEC, SAP, Teradata, and VMware and by the Wallenberg Launch Pad (WALP) grant Dnr 2021.0348. Erik Hellsten and Luigi Nardi were supported by the Wallenberg AI, Autonomous Systems and Software Program (WASP) funded by the Knut and Alice Wallenberg Foundation. Adel Ejjeh was supported in part by funding from an Intel Research Award, from IBM under the DARPA DSSoC program, and from the University of Illinois. Olivia Hsu was supported by an NSF GRFP Fellowship. This research was also supported in part by the Google Research Scholar program. The computations were enabled by resources provided by the Swedish National Infrastructure for Computing (SNIC) at LUNARC, partially funded by the Swedish Research Council through grant agreement no. 2018-05973.

Any opinions, findings, and conclusions or recommendations expressed in this material are those of the authors and do not necessarily reflect the views of the aforementioned funding agencies.

\FloatBarrier

\newpage
\bibliographystyle{plain}
\bibliography{references}

\newpage
\onecolumn
\appendix
\section{Additional Results}
\label{app:additional_results}
In this Appendix, we present the results shown in Sec.~\ref{sec:results} in greater detail.

\begin{figure*}[]
    \centering
    \includegraphics[width = .99\textwidth]{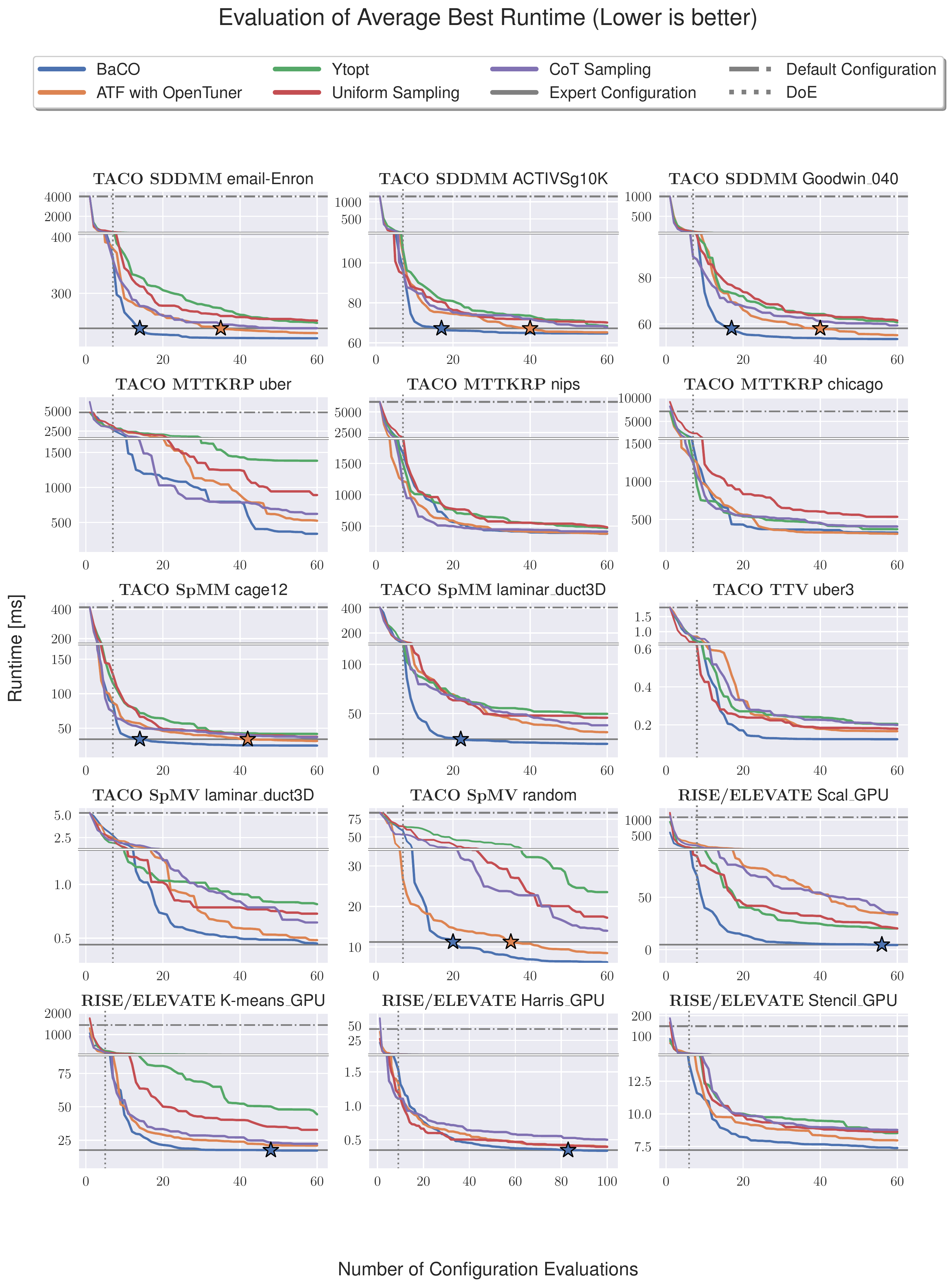}
    \caption{Evolution of average best runtime among evaluated configurations for all benchmarks in Table~\ref{tab:benchmarks} that are not shown in Fig.~\ref{fig:line}. The figure is split vertically into two different scales and the iteration where each method beats the expert configuration is marked with a star.}
    \label{fig:eval:complex}
\end{figure*}

Fig.~\ref{fig:eval:complex} shows the remaining benchmarks not in Fig.~\ref{fig:line}. The takeaway is similar, that BaCO finds the expert-level-performing configurations much faster than the other benchmarks, and that it consistently finds better configurations than the other benchmarks after only a few iterations after the learning phase starts. However, the additional benchmarks indicates that it generalizes well.

\medskip

Table~\ref{tab:expert_level_achieved} shows for each autotuning framework and benchmark how many individual autotuning runs with a full budget managed to reach the expert performance.
The results show that BaCO reaches the expert-level performance in 575 out of 750 runs (76\%).

\medskip

Tables~\ref{tab:relative_perf_33}, \ref{tab:relative_perf_66}, and \ref{tab:relative_perf_100} show the relative performance achieved compared to an expert with the tiny, small, and full budgets. Table~\ref{tab:faster_performance} presents the factors showing how much faster BaCO reaches the best performance of the other methods.
On average, BaCO reaches ATF's performance 2.87$\times$ faster and Ytopt's performance 3.87$\times$ faster.

\FloatBarrier

\begin{table}[tb]
    \centering
    \small
    \begin{tabular}{|c|c|ccccc|}
        \hline
        Framework & Benchmark & BO & ATF & Ytopt & Uniform & CoT\\\hline
        TACO (SpMM) & scircuit & 30& 24& 16& 14& 12\\
        TACO (SpMM) & cage12 & 30& 22& 8& 12& 13\\
        TACO (SpMM) & laminar\-duct3D & 30& 18& 6& 6& 9\\
        TACO (SDDMM) & email-Enron & 30& 25& 17& 10& 18\\
        TACO (SDDMM) & ACTIVSg10K & 30& 28& 19& 11& 14\\
        TACO (SDDMM) & Goodwin\-040 & 30& 24& 18& 10& 12\\
        TACO (MTTKRP) & uber & 24& 21& 4& 3& 6\\
        TACO (MTTKRP) & nips & 24& 20& 9& 7& 15\\
        TACO (MTTKRP) & chicago & 21& 22& 14& 0& 0\\
        TACO (TTV) & facebook & 30& 27& 16& 22& 20\\
        TACO (TTV) & uber3 & 16& 14& 3& 3& 3\\
        TACO (TTV) & random & 18& 17& 16& 17& 17\\
        TACO (SpMV) & laminar\-duct3D & 13& 14& 1& 1& 3\\
        TACO (SpMV) & cage12 & 13& 27& 8& 9& 8\\
        TACO (SpMV) & filter3D & 9& 27& 3& 4& 2\\\hline
        \multicolumn{2}{|c|}{TACO} & 348& 330& 158& 129& 152\\\hline
        RISE \& ELEVATE & MM-CPU & 22& 15& 6& 2& 1\\
        RISE \& ELEVATE & MM-GPU & 19& 10& 1& 1& 0\\
        RISE \& ELEVATE & Asum-GPU & 29& 9& 5& 3& 3\\
        RISE \& ELEVATE & Scal-GPU & 21& 1& 0& 2& 1\\
        RISE \& ELEVATE & K-means-GPU & 25& 15& 3& 5& 6\\
        RISE \& ELEVATE & Stencil-GPU & 18& 4& 8& 0& 0\\\hline
        \multicolumn{2}{|c|}{RISE \& ELEVATE} & 149& 68& 23& 23& 13\\\hline
        HPVM2FPGA & BFS & 29& 0& 4& 2& 7\\
        HPVM2FPGA & Audio & 23& 4& 1& 0& 0\\
        HPVM2FPGA & PreEuler & 26& 6& 4& 6& 4\\\hline
        \multicolumn{2}{|c|}{HPVM2FPGA} & 78& 10& 9& 8& 11\\
        
        \hline
    \end{tabular}
    \caption{Out of 30 autotuning runs with the full budget, how many reached expert-level performance.}
    \label{tab:expert_level_achieved}
\end{table}

\begin{table}[]
    \centering
    \small
    \begin{tabular}{|c|c|ccccc|}
        \hline
        Framework & Benchmark & BaCO & ATF & Ytopt & Uniform & CoT\\\hline        
        TACO (SpMM) & scircuit & 1.16& 0.67& 0.58& 0.59& 0.61\\
        TACO (SpMM) & cage12 & 1.14& 0.74& 0.53& 0.70& 0.72\\
        TACO (SpMM) & laminar\-duct3D & 0.96& 0.34& 0.35& 0.38& 0.35\\
        TACO (SDDMM) & email-Enron & 1.05& 0.91& 0.78& 0.86& 0.91\\
        TACO (SDDMM) & ACTIVSg10K & 1.02& 0.90& 0.83& 0.88& 0.89\\
        TACO (SDDMM) & Goodwin\-040 & 1.04& 0.88& 0.81& 0.79& 0.86\\
        TACO (MTTKRP) & uber & 0.30& 0.20& 0.16& 0.17& 0.33\\
        TACO (MTTKRP) & nips & 0.67& 0.64& 0.50& 0.49& 0.75\\
        TACO (MTTKRP) & chicago & 0.76& 0.58& 0.61& 0.39& 0.60\\
        TACO (TTV) & facebook & 0.70& 0.25& 0.33& 0.52& 0.47\\
        TACO (TTV) & uber3 & 0.73& 0.38& 0.46& 0.51& 0.38\\
        TACO (TTV) & random & 1.30& 0.86& 0.35& 0.92& 1.26\\
        TACO (SpMV) & laminar\-duct3D & 0.60& 0.35& 0.42& 0.43& 0.36\\
        TACO (SpMV) & cage12 & 0.53& 0.85& 0.33& 0.30& 0.39\\\hline
        \multicolumn{2}{|c|}{TACO}& 0.83& 0.63& 0.49& 0.54& 0.61\\\hline
        RISE \& ELEVATE & MM-CPU & 0.89& 0.81& 0.79& 0.78& 0.78\\
        RISE \& ELEVATE & MM-GPU & 0.28& 0.17& 0.28& 0.25& 0.14\\
        RISE \& ELEVATE & Asum-GPU & 0.59& 0.35& 0.34& 0.29& 0.35\\
        RISE \& ELEVATE & Scal-GPU & 0.34& 0.06& 0.12& 0.11& 0.06\\
        RISE \& ELEVATE & K-means-GPU & 0.82& 0.61& 0.22& 0.35& 0.53\\
        RISE \& ELEVATE & Stencil-GPU & 0.88& 0.78& 0.72& 0.73& 0.73\\\hline
        \multicolumn{2}{|c|}{RISE \& ELEVATE}& 0.65& 0.48& 0.41& 0.46& 0.44\\\hline
        HPVM2FPGA & BFS & 0.48& 0.18& 0.60& 0.28& 0.29\\
        HPVM2FPGA & Audio & 0.93& 0.71& 0.83& 0.64& 0.70\\
        HPVM2FPGA & PreEuler & 0.46& 0.36& 0.32& 0.37& 0.38\\\hline
        \multicolumn{2}{|c|}{HPVM2FPGA}& 0.62& 0.42& 0.58& 0.43& 0.46\\\hline
        \multicolumn{2}{|c|}{All} & 0.76 & 0.56 & 0.49 & 0.50 & 0.55\\
        \hline
    \end{tabular}
    \caption{Relative performance compared to expert with the tiny budget (1/3 of the full budget).
    Values larger than 1 indicate a performance advantage over the expert.
    Values below 1 indicate a performance disadvantage compared with the expert.
    }
    \label{tab:relative_perf_33}
\end{table}

\begin{table}[]
    \centering
    \small
    \begin{tabular}{|c|c|ccccc|}
        \hline
        Framework & Benchmark & BaCO & ATF & Ytopt & Uniform & CoT\\\hline
        TACO (SpMM) & scircuit & 1.25& 0.96& 0.77& 0.78& 0.77\\
        TACO (SpMM) & cage12 & 1.35& 0.99& 0.80& 0.84& 0.82\\
        TACO (SpMM) & laminar\-duct3D & 1.16& 0.60& 0.45& 0.50& 0.51\\
        TACO (SDDMM) & email-Enron & 1.08& 1.01& 0.91& 0.92& 0.98\\
        TACO (SDDMM) & ACTIVSg10K & 1.04& 1.01& 0.91& 0.94& 0.93\\
        TACO (SDDMM) & Goodwin\-040 & 1.08& 1.00& 0.90& 0.91& 0.95\\
        TACO (MTTKRP) & uber & 0.42& 0.38& 0.24& 0.27& 0.43\\
        TACO (MTTKRP) & nips & 0.94& 0.92& 0.68& 0.68& 0.85\\
        TACO (MTTKRP) & chicago & 0.90& 1.00& 0.74& 0.54& 0.72\\
        TACO (TTV) & facebook & 2.13& 0.75& 0.53& 0.93& 0.87\\
        TACO (TTV) & uber3 & 1.00& 0.69& 0.52& 0.67& 0.55\\
        TACO (TTV) & random & 6.29& 1.42& 0.84& 1.83& 2.40\\
        TACO (SpMV) & laminar\-duct3D & 0.88& 0.74& 0.48& 0.56& 0.52\\
        TACO (SpMV) & cage12 & 0.65& 1.17& 0.44& 0.48& 0.54\\
        TACO (SpMV) & filter3D & 0.66& 1.28& 0.42& 0.34& 0.43\\\hline
        \multicolumn{2}{|c|}{TACO} & 1.39& 0.93& 0.64& 0.75& 0.82\\\hline
        RISE \& ELEVATE & MM-CPU & 0.98& 0.90& 0.85& 0.81& 0.82\\
        RISE \& ELEVATE & MM-GPU & 0.84& 0.38& 0.40& 0.36& 0.19\\
        RISE \& ELEVATE & Asum-GPU & 1.20& 0.58& 0.38& 0.41& 0.49\\
        RISE \& ELEVATE & Scal-GPU & 0.88& 0.09& 0.19& 0.15& 0.09\\
        RISE \& ELEVATE & K-means-GPU & 1.00& 0.74& 0.34& 0.44& 0.65\\
        RISE \& ELEVATE & Stencil-GPU & 0.94& 0.86& 0.76& 0.81& 0.80\\
        RISE \& ELEVATE & Harris-GPU & 0.95& 0.79& 0.80& 0.61\\\hline
        \multicolumn{2}{|c|}{RISE \& ELEVATE} & 0.97& 0.62& 0.49& 0.54& 0.52\\\hline
        HPVM2FPGA & BFS & 1.00& 0.34& 0.70& 0.56& 0.50\\
        HPVM2FPGA & Audio & 1.00& 0.90& 0.91& 0.86& 0.83\\
        HPVM2FPGA & PreEuler & 0.78& 0.45& 0.36& 0.50& 0.48\\\hline
        \multicolumn{2}{|c|}{HPVM2FPGA} & 0.92& 0.57& 0.66& 0.64& 0.61\\\hline
        \multicolumn{2}{|c|}{All} & 1.22 & 0.80 & 0.61 & 0.67 & 0.71\\
        
        \hline
    \end{tabular}
    \caption{Relative performance compared to expert with the small budget (2/3 of the full budget).
    Values larger than 1 indicate a performance advantage over the expert.
    Values below 1 indicate a performance disadvantage compared with the expert.
    }
    \label{tab:relative_perf_66}
\end{table}

\begin{table}[]
    \centering
    \small
    \begin{tabular}{|c|c|ccccc|}
        \hline
        Framework & Benchmark & BaCO & ATF & Ytopt & Uniform & CoT\\\hline
        TACO (SpMM) & scircuit & 1.31& 1.12& 0.85& 0.90& 0.91\\
        TACO (SpMM) & cage12 & 1.37& 1.08& 0.82& 0.90& 0.92\\
        TACO (SpMM) & laminar\-duct3D & 1.24& 0.77& 0.48& 0.52& 0.63\\
        TACO (SDDMM) & email-Enron & 1.08& 1.04& 0.96& 0.95& 1.00\\
        TACO (SDDMM) & ACTIVSg10K & 1.04& 1.03& 0.98& 0.96& 0.99\\
        TACO (SDDMM) & Goodwin\-040 & 1.09& 1.05& 0.96& 0.95& 0.98\\
        TACO (MTTKRP) & uber & 1.00& 0.64& 0.24& 0.38& 0.54\\
        TACO (MTTKRP) & nips & 1.00& 1.00& 0.81& 0.78& 0.91\\
        TACO (MTTKRP) & chicago & 1.00& 1.05& 0.88& 0.61& 0.80\\
        TACO (TTV) & facebook & 2.52& 1.90& 0.65& 1.25& 1.16\\
        TACO (TTV) & uber3 & 1.00& 0.75& 0.61& 0.69& 0.62\\
        TACO (TTV) & random & 8.82& 3.32& 1.81& 2.32& 2.90\\
        TACO (SpMV) & laminar\-duct3D & 0.98& 0.91& 0.54& 0.60& 0.68\\
        TACO (SpMV) & cage12 & 0.77& 1.48& 0.47& 0.63& 0.62\\
        \hline\multicolumn{2}{|c|}{TACO} & 1.66& 1.24& 0.77& 0.86& 0.95\\\hline
        RISE \& ELEVATE & MM-CPU & 1.02& 0.93& 0.88& 0.83& 0.83\\
        RISE \& ELEVATE & MM-GPU & 1.05& 0.49& 0.49& 0.44& 0.29\\
        RISE \& ELEVATE & Asum-GPU & 1.25& 0.67& 0.43& 0.50& 0.58\\
        RISE \& ELEVATE & Scal-GPU & 1.08& 0.14& 0.23& 0.23& 0.14\\
        RISE \& ELEVATE & K-means-GPU & 1.02& 0.84& 0.40& 0.54& 0.79\\
        RISE \& ELEVATE & Stencil-GPU & 0.98& 0.91& 0.85& 0.84& 0.82\\
        \hline\multicolumn{2}{|c|}{RISE \& ELEVATE} & 1.06& 0.69& 0.55& 0.61& 0.59\\\hline
        HPVM2FPGA & BFS & 1.00& 0.45& 0.74& 0.69& 0.67\\
        HPVM2FPGA & Audio & 1.00& 0.94& 0.94& 0.90& 0.91\\
        HPVM2FPGA & PreEuler & 1.00& 0.50& 0.51& 0.57& 0.56\\
        \hline\multicolumn{2}{|c|}{HPVM2FPGA} & 1.00& 0.63& 0.73& 0.72& 0.71\\\hline
        \multicolumn{2}{|c|}{All} & 1.41 & 1.01 & 0.71 & 0.76 & 0.83\\        
        \hline
    \end{tabular}
    \caption{Relative performance compared to expert with the full budget.
    Values larger than 1 indicate a performance advantage over the expert.
    Values below 1 indicate a performance disadvantage compared with the expert.
    }
    \label{tab:relative_perf_100}
\end{table}

\begin{table}[]
    \centering
    \small
    \begin{tabular}{|c|c|cccc|}
        \hline
        Framework & Benchmark & ATF & Ytopt & Uniform & CoT\\\hline
        TACO (SpMM) & scircuit & 3.33$\times$& 5.00$\times$& 4.62$\times$& 4.62$\times$\\
        TACO (SpMM) & cage12 & 3.53$\times$& 6.00$\times$& 5.45$\times$& 5.00$\times$\\
        TACO (SpMM) & laminar\-duct3D & 4.00$\times$& 5.45$\times$& 5.00$\times$& 4.29$\times$\\
        TACO (SDDMM) & email-Enron & 3.75$\times$& 4.62$\times$& 4.62$\times$& 4.29$\times$\\
        TACO (SDDMM) & ACTIVSg10K & 1.88$\times$& 4.62$\times$& 5.45$\times$& 3.75$\times$\\
        TACO (SDDMM) & Goodwin\-040 & 2.40$\times$& 4.29$\times$& 4.62$\times$& 4.00$\times$\\
        TACO (MTTKRP) & uber & 1.40$\times$& 4.62$\times$& 1.88$\times$& 1.43$\times$\\
        TACO (MTTKRP) & nips & 1.03$\times$& 2.31$\times$& 2.40$\times$& 1.58$\times$\\
        TACO (MTTKRP) & chicago & - & 2.14$\times$& 3.53$\times$& 2.86$\times$\\
        TACO (TTV) & facebook & 1.76$\times$& 3.00$\times$& 2.40$\times$& 2.50$\times$\\
        TACO (TTV) & uber3 & 2.86$\times$& 3.53$\times$& 3.16$\times$& 3.33$\times$\\
        TACO (TTV) & random & 1.67$\times$& 2.07$\times$& 2.07$\times$& 1.67$\times$\\
        TACO (SpMV) & laminar\-duct3D & 1.09$\times$& 3.33$\times$& 2.86$\times$& 2.73$\times$\\
        TACO (SpMV) & cage12 & - & 4.00$\times$& 1.76$\times$& 1.76$\times$\\
        \hline\multicolumn{2}{|c|}{TACO} &3.15$\times$&5.00$\times$&4.96$\times$&4.32$\times$\\\hline
        RISE \& ELEVATE & MM-CPU & 2.22$\times$& 3.03$\times$& 3.85$\times$& 3.85$\times$\\
        RISE \& ELEVATE & MM-GPU & 2.03$\times$& 1.97$\times$& 2.18$\times$& 2.93$\times$\\
        RISE \& ELEVATE & Asum-GPU & 2.73$\times$& 3.53$\times$& 3.33$\times$& 3.00$\times$\\
        RISE \& ELEVATE & Scal-GPU & 4.62$\times$& 3.75$\times$& 3.75$\times$& 4.62$\times$\\
        RISE \& ELEVATE & K-means-GPU & 2.73$\times$& 6.00$\times$& 5.00$\times$& 3.16$\times$\\
        RISE \& ELEVATE & Stencil-GPU & 2.50$\times$& 3.53$\times$& 3.53$\times$& 3.75$\times$\\
        \hline\multicolumn{2}{|c|}{RISE \& ELEVATE} &2.68$\times$&3.38$\times$&3.51$\times$&3.55$\times$\\\hline
        HPVM2FPGA & BFS & 2.86$\times$& 2.00$\times$& 2.22$\times$& 2.22$\times$\\
        HPVM2FPGA & Audio & 2.86$\times$& 2.86$\times$& 3.16$\times$& 3.00$\times$\\
        HPMV2FPGA & PreEuler & 2.61$\times$& 2.61$\times$& 2.00$\times$& 2.07$\times$\\
        \hline\multicolumn{2}{|c|}{HPVM2FPGA} &2.77$\times$&2.49$\times$&2.46$\times$&2.43$\times$\\\hline
        \multicolumn{2}{|c|}{all} &2.87$\times$&3.82$\times$&3.86$\times$&3.63$\times$\\
        
        \hline
    \end{tabular}
    \caption{
    Factors showing how much faster BaCO reach the best performance of the other methods.
    E.g., a factor of 3.33$\times$ indicates that BaCO required 3.33$\times$ less evaluations to achieve the same performance. ''-'' indicates that the average final performance of BaCO was lower than ATF.
    }
    \label{tab:faster_performance}
\end{table}

\FloatBarrier

\clearpage

\section{Wall Clock Time Analysis}
\label{sec:wallclock}
In Table~\ref{tab:wallclocktime}, we show the average wall clock time of BaCO and the baselines for the TACO  SpMM and SDDMM benchmarks. 
The breakdown of the wall clock time is given by two main factors: 1) the evaluations of the black-box function, and 2) the computing time for the autotuner to generate its recommendations. 
The most expensive part of the autotuning process is the first factor, where the program kernel is evaluated a number of times corresponding to the budget allocated. Thus, methods that recommend slow-to-evaluate configurations tend to use more wall clock time per evaluation. 
The second factor relates to the internal workings of an autotuner search method. In this regard, the more intricate model-based methods tend to be slower, but this cost becomes less prominent when optimizing larger program kernels. 
This factor is highly dependent on the quality of the software implementation of the autotuner. 
It is beyond the scope of this work to provide the fastest implementation of both BaCO and the baselines — In Table~\ref{tab:wallclocktime} we give a rough analysis of the current average wall clock time for all autotuners, which will likely be improved in future releases of these tools. 
We observe that while BaCO uses a more complex method than the baselines it is the second fastest method behind ATF. 
ATF uses OpenTuner which employs search algorithms based on heuristics — These are usually faster than model-based approaches such as BaCO. 
We can see this insight reflected also in the Ytopt wall clock time. 

\begin{table}[h!]
    \centering
\begin{tabular}{c|ccccc}
     & BaCO & ATF w. & Ytopt & Uniform & CoT  \\
     & & OpenTuner &  & sampling & samp.  \\\hline
    SpMM & 262 & \textbf{144} & 309 & 402 & 336\\
    SDDMM & 263 & \textbf{197} & 274 & 433 & 381
\end{tabular}
    \caption{Average wall clock time in seconds for BaCO and the baselines on the TACO SpMM and SDDMM benchmarks.}
    \label{tab:wallclocktime}
\end{table}

\end{document}